\documentclass{emulateapj}

\newcommand{\ie}{{\it i.e.}}
\newcommand{\eg}{{\it e.g.}}

\def\asec{\hbox{$^{\prime\prime}$}}

\def\micro{$\mu$}
\def\err{$\pm$}

\def\sima{$\sim$}

\def\mina{\hbox{$^\prime$}}
\def\ella{$\ell$}
\def\dega{$^\circ$}
\def\sqra{$^2$}
\def\sims{$\sim$\ }

\def\approxs{$\approx$\ }
\def\mins{\hbox{$^\prime$}\ }

\def\ells{$\ell$\ }
\def\degs{$^\circ$\ }

\def\ditto{\tt "}

\def\mum{$\mathrm{\mu m}\;$}
\def\sin{silicon-nitride}
\def\NFL{$NF\lambda$}

\def\nvrthz{$nV~Hz^{-0.5}$}
\def\Wrthz{$W~Hz^{-0.5}$}
\def\4he{$^4$He}
\def\3he{$^3$He}
\def\n2{N$_2$}
\def\bl{$B_{\ell}$}
\def\cl{$C_{\ell}$}

\def\max{{\sc max}}
\def\wmap{{\sc wmap}}
\def\maxima{{\sc maxima}}

\def\maximai{{\sc maxima-i}}
\def\maximaii{{\sc maxima-ii}}

\begin{document}

\title{MAXIMA: a balloon-borne Cosmic Microwave Background anisotropy experiment}

\author {B.~Rabii\altaffilmark{1,2,3},
  C.~D.~Winant\altaffilmark{1,2,3},
  M.~E.~Abroe\altaffilmark{4},
  P.~Ade\altaffilmark{5},
  A.~Balbi\altaffilmark{6},
  J.~J.~Bock\altaffilmark{7,8},
  J.~Borrill\altaffilmark{2,9},
  A.~Boscaleri\altaffilmark{10},
  P.~de~Bernardis\altaffilmark{11},
  J.~S.~Collins\altaffilmark{1,2},
  P.~G.~Ferreira\altaffilmark{12},
  S.~Hanany\altaffilmark{3,4},
  V.~V.~Hristov\altaffilmark{7},
  A.~H.~Jaffe\altaffilmark{3,13},
  B.~R.~Johnson\altaffilmark{4},
  A.~E.~Lange\altaffilmark{7},
  A.~T.~Lee\altaffilmark{1,2,3},
  C.~B.~Netterfield\altaffilmark{14},
  E.~Pascale\altaffilmark{10},
  P.~L.~Richards\altaffilmark{1,3},
  G.~F.~Smoot\altaffilmark{1,15},
  R.~Stompor\altaffilmark{2,9},
  J.~H.~P.~Wu\altaffilmark{3,16}
}

\altaffiltext{1}{Dept. of Physics, University of California, Berkeley, CA 94720, USA}

\altaffiltext{2}{Space Sciences Laboratory, University of California,
  Berkeley, CA 94720, USA}

\altaffiltext{3}{Center for Particle Astrophysics, University of
  California, Berkeley, CA 94720, USA}

\altaffiltext{4}{School of Physics and Astronomy, University of
  Minnesota/Twin Cities, Minneapolis, MN 55455, USA}

\altaffiltext{5}{Cardiff University, Physics Department, Cardiff CF24 3YB, UK}

\altaffiltext{6}{Universit\'a di Roma Tor Vergata, 00133 Roma, Italy}

\altaffiltext{7}{California Institute of Technology, Pasadena, CA 91125, USA}

\altaffiltext{8}{Jet Propulsion Laboratory, Pasadena, CA 91109, USA}

\altaffiltext{9}{Computational Research Division, Lawrence Berkeley National Laboratory, Berkeley, CA 94720, USA}

\altaffiltext{10}{IFAC-CNR, 50127 Firenze, Italy}

\altaffiltext{11}{Universit\'a di Roma La Sapienza, I-00185 Roma, Italy}

\altaffiltext{12}{Astrophysics and Theoretical Physics, University of Oxford, Oxford OX1 3RH, UK}

\altaffiltext{13}{Imperial College, London SW7 2BW, UK}

\altaffiltext{14}{Physics Department, University of Toronto, Toronto, ON M5S 3H8, Canada}

\altaffiltext{15}{Physics Division, Lawrence Berkeley National Laboratory, Berkeley, CA 94720, USA}

\altaffiltext{16}{Dept. of Physics, National Taiwan University, Taipei 106, Taiwan}


\begin{abstract}

We describe the Millimeter wave Anisotropy eXperiment IMaging Array
(\maxima), a balloon-borne experiment designed to measure the
temperature anisotropy of the Cosmic Microwave Background (CMB) on
angular scales of 10\mins to 5\dega .  \maxima\ mapped the CMB using
16 bolometric detectors in spectral bands centered at 150~GHz,
240~GHz, and 410~GHz, with 10\mins resolution at all frequencies.  The
combined receiver sensitivity to CMB anisotropy was \sima 40~\micro
K~$\sqrt{sec}$.  Systematic parasitic contributions were minimized by
using four uncorrelated spatial modulations, thorough crosslinking,
multiple independent CMB observations, heavily baffled optics, and
strong spectral discrimination.  Pointing reconstruction was accurate
to 1\mins, and absolute calibration was better than 4\%.  Two \maxima\
flights with more than 8.5 hours of CMB observations have mapped a
total of 300~deg$^2$ of the sky in regions of negligible known
foreground emission. \maxima\ results have been released in previous
publications. \maximai\ maps, power spectra and correlation matrices
are publicly available at http://cosmology.berkeley.edu/maxima.

\end{abstract}




\section{INTRODUCTION}\label{introduction}

        \maxima\ was a balloon-borne experiment designed to measure
the anisotropy of the CMB over a wide range of angular scales ($\ell
=$ 35 to 1000).  High resolution observations were made of 300~deg$^2$
of the sky over the course of two flights, in 1998 and 1999.  Results
have been released (\citet{lee_results}, \citet{hanany_results}) and
cosmological implications have been explored from both the \maxima\ data
set alone (\citet{stompor_results}, \citet{abroe_results},
\citet{balbi_results}, \citet{cayon}) and combined with other
data sets (\citet{maxiboom}, \citet{maxicbi}, \citet{maxivsa},
\citet{maxiacbar}). \citet{abroe_crosscor} presents correlations of
\maxima\ data with those from the Wilkinson Microwave Anisotropy Probe
\citep{wmap}.  

This paper, which is derived from two Ph.D. dissertations
(\citet{BRThesis}, \citet{CDWThesis}), is an overview of the
experimental design and achieved performance of \maxima.
Section~\ref{observations} describes the two \maxima\ science
flights. Section~\ref{optics} describes the optics and optical
characterization.  Section~\ref{detectors} describes the detector
system.  Section~\ref{receiver} describes the cryogenic receiver and
support electronics.  Section~\ref{calibration} discusses the
in-flight responsivity calibration.  Section~\ref{pointing} presents
the pointing system and attitude reconstruction.

\subsection{Goals}

The primary scientific objectives of \maxima\ were to distinguish
between the inflationary paradigm and the topological defect models
for the evolution of the Universe, to measure the geometry of the
Universe by identifying the location of the first peak in the CMB
power spectrum and to provide information about other cosmological
parameters.  The $\ell$-space coverage and resolution of the
experiment were well suited for measurement of the first three
acoustic peaks of adiabatic inflationary models.  Measurements in this
region have been a powerful tool for testing the general predictions
of inflation and for parameter estimation.  In the past few years, a
number of experiments have published significant measurements of this
type (\citet{wmap}, \citet{wang}).

        \maxima\ data have also been used to test analysis methods and
tools.  Treatments have been developed for problems such as beam
asymmetry \citep{wu_beam}, foreground discrimination
\citep{jaffe_dust}, scan synchronous noise
\citep{stompor_mapmaking}), and detection of spatial non-Gaussianity
(\citet{wu_gauss}, \citet{bispectrum}, \citet{bispectrum2}).

        \maxima\ has been used to test new technologies.  In
particular, \maxima\ was the first CMB experiment to have used \sima
100-mK spider-web bolometers, similar to those planned for the Planck
Surveyor.  The combination of these detectors and an adiabatic
demagnetization cooling system has provided receiver sensitivity of
\sima 40~\micro K~$\sqrt{sec}$, the best reported by any CMB
experiment.

\begin{figure}[ht]
\plotone{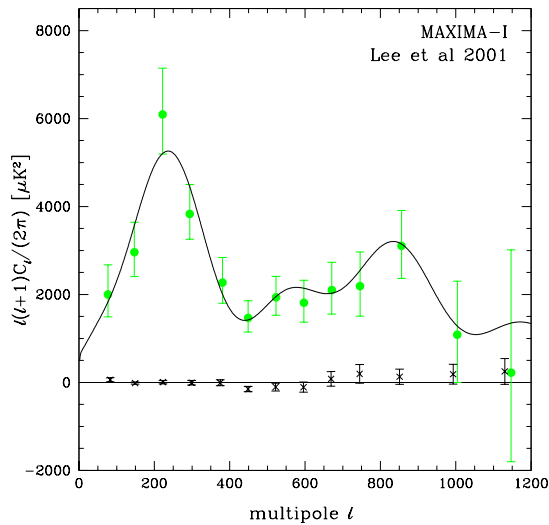}
\caption{The power spectrum of the
CMB using a hybrid analysis of 5\mins resolution (up to \ells = 335)
and 3\mins resolution (over \ells = 335) maps.  Error bars show the
statistical uncertainties.  The solid curve is the power spectrum of
the best fit model from \citet{balbi_results} with $\Omega_b = 0.1$,
$\Omega_{cdm} = 0.6$, $\Omega_{\Lambda} = 0.3$, $n = 1.08$, and $h =
0.53$.  The crosses are the power spectrum of the difference between
the map from one detector and the combined map from the other two
detectors used for the 3\mins analysis.  (\citet{lee_results})}
\label{fig:cmbps}
\end{figure}

The primary scientific result from \maxima\ is the angular power
spectrum shown in Figure~\ref{fig:cmbps} \citep{lee_results}.

\subsection{Technical Overview}

\begin{figure}[ht]
\plotone{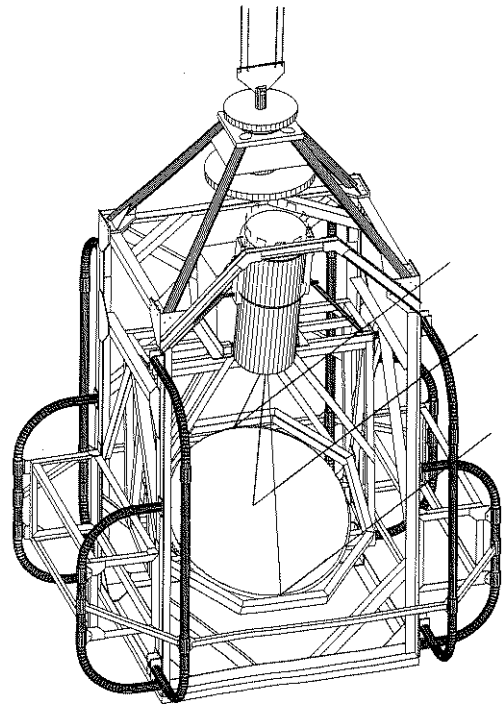}
\caption{A mechanical drawing of the \maxima\
telescope from an elevated front/side perspective.  Rays representing
the telescope beam are shown reflecting from the primary mirror into
the cryogenic receiver.  Electronics housed in the rectangular boxes
on the sides of the instrument include the pointing system, data
multiplexers and digitizers, and telemetry and command interfaces.
Near the top of the telescope are motors controlling azimuthal
orientation.  The inner assembly consisting of the primary mirror and
the receiver is tilted relative to the outer frame to aim the
telescope in elevation.  The gondola frame is covered in lightweight
aluminum-covered builders foam (not shown) and the primary mirror is
surrounded by a scoop built of aluminum sheet (not shown), both of
which shield the telescope and receiver from stray optical and radio
frequency radiation.}
\label{fig:wireframe}
\end{figure}

        \maxima\ was a bolometric instrument which measured CMB
temperature fluctuations in frequency bands centered at 150~GHz,
240~GHz, and 410~GHz.  In order to avoid atmospheric emission,
observations were made from an altitude of \sima 40~km during multiple
balloon flights.  The relatively short duration of the balloon-borne
observations was offset by the use of a 16-element array of single
color photometers with extremely sensitive detectors.  The telescope
was an off-axis Gregorian system with a 1.3-m diameter primary mirror
providing a 10\mins beam size (FWHM) for all detectors.  The
combination of this angular resolution and $\sim 100$ square degrees
of sky coverage made the experiment sensitive over a wide range of
angular scales.  The compact and well cross-linked scan pattern was
optimized for extracting the angular power spectrum.  The use of three
spectral bands allowed discrimination between the CMB and foreground
sources.
\maxima\ benefited from precise pointing reconstruction (1\mins rms)
and accurate calibration (4\%).  The instrument was designed to
survive repeated balloon flights and has been successfully recovered
after five flights, including two \maxima\ science flights.  A
mechanical drawing of the \maxima\ telescope can be seen in
Figure~\ref{fig:wireframe}.

\section{OBSERVATIONS}\label{observations}

		The first science flight, \maximai, was launched on
1998 August 2 at 00:58 UT (1998 August 1, 19:58 local time) from the
National Scientific Balloon Facility (NSBF) in Palestine, TX (latitude
31.8\dega N, longitude 95.7\dega W).  The maximum float altitude of
37.5~km was reached at 4:35 UT.  The telescope traveled 189~km west
and less than 1~km south before reaching maximum altitude.  At float,
the telescope drifted 405~km west and less than 1~km north.  Descent
began 3.8~hours later at 8:22 UT.  Summer flights from the NSBF in
Palestine are limited to a range of approximately 600~km.  The
\maximai\ flight was relatively short due to fast high altitude winds.

\begin{figure}[ht]
\plotone{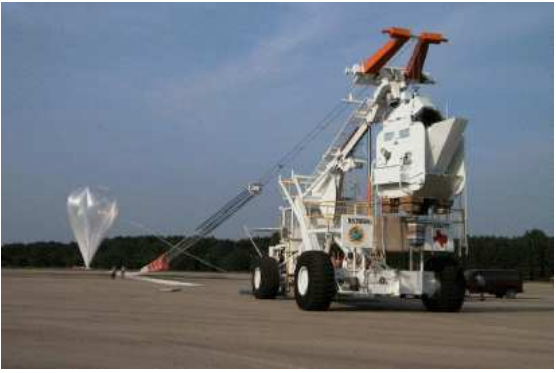}
\caption{The \maxima\ telescope on the launch vehicle on 1999 June 16}
\label{fig:launch}
\end{figure}

		Four observations were conducted during the flight.
First, the CMB dipole was observed in order to calibrate the
responsivity of the detectors.  The dipole observation was started
before reaching float altitude and lasted from 03:37 UT to 04:11 UT.
Next, two overlapping, cross-linked scans of CMB anisotropy were
conducted over a 122-deg$^2$ region in the vicinity of the Draconis
constellation.  These scans occurred from 04:21 UT to 05:59 UT and
06:02 UT to 07:24 UT.  Finally, observations were made of Jupiter to
characterize the telescope beams and to calibrate the 410-GHz
detectors, which were insensitive to the CMB dipole.  Jupiter was
observed from 07:30 UT until 08:04 UT.

	The Sun was at least 20\degs below the horizon for all
observations.  The Moon was below 20\degs elevation during CMB
observations, and below the horizon for over an hour of the second
observation.  While above the horizon, it was at least 70\degs from
the scan region.  The relative position of the Moon differed by
20\degs azimuth and 10\degs elevation between the two CMB scans.
During the dipole observation, the Moon was at 30\degs elevation,
20\degs below the scan.  The Moon was below the horizon during the
Jupiter scan.  The Moon was 68\% full during the flight.

	The instrument was launched a second time (\maximaii) on 1999 June
17 at 00:07 UT (1999 June 16, 19:07 local time).
Figure~\ref{fig:launch} is a photograph taken shortly before the
launch.  The telescope traveled 42~km east and 9~km south before
reversing direction and reaching maximum altitude 97~km west and 1~km
south of the launch position.  The maximum float altitude of 38.0~km
was reached at 04:34 UT.  At float, the telescope drifted 490~km west
and 42~km north.  Descent began 7.8~hours later at 12:21 UT.  The
relatively slow high-altitude winds of early summer allowed us a
considerably longer flight than
\maximai.

	As with \maximai, two CMB observations and two calibration
scans were conducted.  The first was an observation of Mars from 03:14
UT to 03:52 UT.  Approximately one hour was spent on maintenance
tasks.  Two overlapping, cross-linked CMB scans were conducted from
05:04 UT to 07:29 UT and 07:31 UT to 09:40 UT.  The observed region
had an area of 225~deg$^2$ and overlapped the \maximai\ region by
50~deg$^2$.  A calibration scan of the CMB Dipole was conducted from
09:42 UT to 10:19 UT.  Further data were recorded from 10:20 UT to
11:59 UT as a test of the daytime performance of the instrument.

	The Sun rose to -20\degs elevation at 09:24 UT and to 0\degs
elevation at 11:17 UT.  Data collected after 10:20 UT have been used
only as test data for future daytime balloon flights.  The Moon was
17\% full during the flight and was below the horizon during the
dipole observation and both CMB observations.  During the Mars
observation, the Moon was 75\degs from the scan.

	The scans for both flights are summarized in
Table~\ref{tab:flight_stats}.

\begin{table}
\caption{Flight Statistics}
\begin{center}
\begin{tabular}{ccccccc}
\hline
Flight & Hrs at          & 1st CMB     & 2nd CMB     & CMB         & Planet   & Daytime
\\     	 & Max             & Scan        & Scan        & Dipole      & Scan   	& Test Data
\\       & Alt 	           & (hrs) 	 & (hrs)       & (hrs) 	     & (hrs) 	& (hrs)
\\\hline
\maximai & 3.78$^{a}$ & 1.63 & 1.37 & 0.57 & 0.57 & 0.00 
\\\maximaii & 7.78 & 2.42 & 2.15 & 0.62 & 0.63 & 1.65 
\\\hline
\multicolumn{7}{l}{$^{a}$ Some calibration data were collected before the}\\
\multicolumn{7}{l}{telescope reached maximum altitude, resulting in a total}\\
\multicolumn{7}{l}{scan time greater than the time at maximum altitude.}
\end{tabular}
\label{tab:flight_stats}
\end{center}
\end{table}

\subsection{Sky Selection}

\begin{figure}[ht]
\plotone{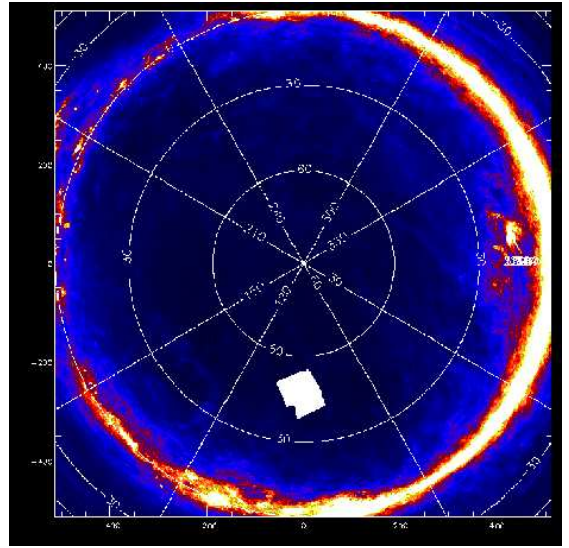}
\caption{The \maximai\ CMB observation
scan (white) is plotted over the Berkeley-Durham IRAS-Dirbe map of the
northern galactic hemisphere dust emission extrapolated to 150~GHz
\citep{fds_dust}.  Areas of the sky with low dust contrast are shown
as dark regions of the map. The \maximai\ CMB observation region is
constrained to $b > 30^{\circ}$.}
\label{fig:m1dirbe}
\end{figure}

	The main constraint on sky selection has been celestial
foregrounds.  The scan regions have a predicted dust temperature
anisotropy of \sima 10.0~\micro K at 150~GHz with rms fluctuations of
\sima 2.5~\micro K in units of equivalent CMB temperature fluctuation
\citep{forecast}.  Tests of the spectral and angular profiles of the
observed signals, as well as cross correlations with known dust maps,
confirmed the absence of significant dust contamination in our CMB
data \citep{jaffe_dust}.  The \maximai\ scan region was chosen to
contain no detectable point sources.  For \maximaii, this requirement
was relaxed so that a few bright sources might be detectable in the
anisotropy maps, particularly at 410~GHz.  No point source
contribution is expected in the CMB power spectrum.  The \maximai\
scan region is shown overlaid with the IRAS-Dirbe dust map in
Figure~\ref{fig:m1dirbe}.

\begin{figure}[ht]
\plotone{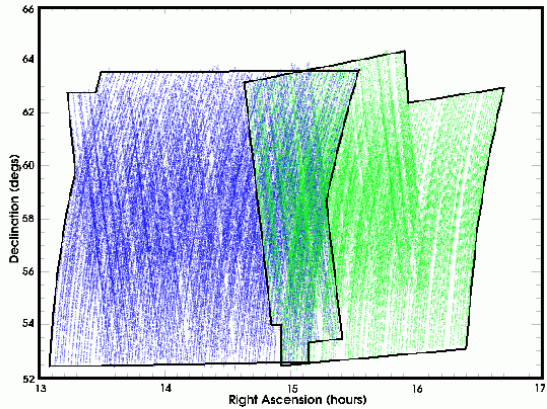}
\caption{The full reconstructed pointing for a single
detector in both \maxima\ flights.  \maximai\ is the lighter region on
the right, and \maximaii\ is the darker region on the left.  The scan
region for each flight is boxed, and the \sima 50-deg$^2$ overlap
region can be seen at Right Ascension \sima 15~hours.}
\label{fig:colorscans}
\end{figure}

		The scan regions for the two flights were chosen to
have a modest (\sima 50-deg$^2$) overlap, both as a consistency check
and to facilitate the combination of the data sets.  The combined scan
pattern is shown in Figure~\ref{fig:colorscans}.

\section{OPTICS}\label{optics}

In order to achieve the desired sensitivity to CMB temperature
fluctuations up to multipoles of $\ell \simeq 1000$, the telescope had
an angular resolution of 10\mins FWHM.  To make sure that local
sources such as the Earth, the Moon and the balloon did not contaminate
the measurement, we rejected side-lobe signals with a heavily baffled
optical system.  Optical background loading was on order of that from
the CMB itself.  Spectral bands were defined to distinguish CMB from
atmosphere and celestial foregrounds.

\subsection{Optical Design} \label{optsys}

\subsubsection{The Telescope}

\begin{figure*}[ht]
\plotone{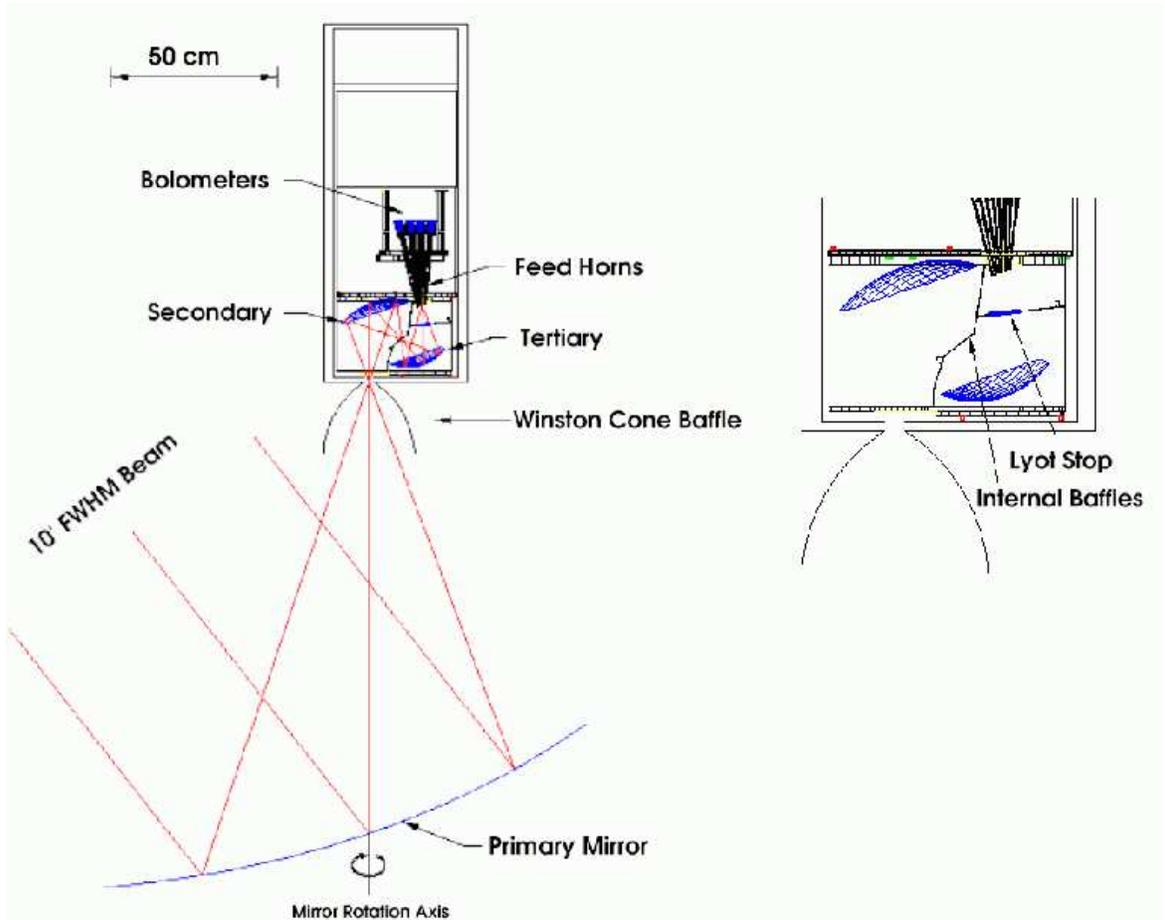}
\caption{{\bf{Left:}} The telescope was a fast (f/1) Gregorian
system.  The primary mirror, a 1.3-m diameter underilluminated
paraboloid, was modulated about the indicated axis.  A Winston cone
baffle (shown outside the cryostat window) blocked radiation not
reflected from the primary mirror.  Cooled secondary and tertiary
mirrors corrected aberrations from the primary.  An array of feed
horns channeled light to the bolometer cavities.  Optical filters were
located at the prime focus, the Lyot stop, and after the feed horns.
{\bf{Right:}} The focal plane was baffled from stray light in several
ways.  At least 5 edge diffractions with angles of up to 65 degrees
were required for rays outside of the defined throughput to arrive to
the focal plane.  The internal baffles were blackened with mm
absorptive material and the Winston cone baffle restricted the
throughput to the area of the primary mirror. A Lyot stop defined the
illumination on the primary mirror.}
\label{fig:optics}
\end{figure*}

The \maxima\ telescope was an off-axis Gregorian system, consisting of a
\sima 1.3 m diameter underfilled primary mirror with cooled, baffled
ellipsoidal secondary and tertiary mirrors.  
Figure~\ref{fig:optics} shows the optical system and
Table~\ref{tab:optics_specs} summarizes its optical properties.

\begin{table}
\caption{Optical Parameters of the MAXIMA Telescope.}
\begin{center}
\begin{tabular}{lr}
\hline \hline
{\bf System Parameters} & \\
Effective Pupil Diameter        & 835 mm\\
System Focal Length             & 1521 mm \\
Focal Ratio                     & 1.8 \\
Plate Scale                     & 0.4 deg~cm$^{-1}$ \\
\hline
{\bf Primary Mirror} & \\
Dimensions                      & 1328 mm $\times$ 1267 mm \\
Conic Constant$^{a}$            & -1 \\
Vertex Radius of Curvature      & 2400 mm \\
Focal Length                    & 1340 mm\\
\hline
{\bf Secondary Mirror} & \\
Radius                          & 210 mm \\
Conic Constant                  & -0.126 \\
Vertex Radius of Curvature      & 168 mm\\
Aspheric coefficients$^{a}$           & $A=2.15\times10^{-5}$ \\
                                & $B=-9.07\times10^{-8}$ \\
                                & $C=1.55\times10^{-10}$  \\
\hline
{\bf Tertiary mirror} & \\
Radius                          & 180 mm \\
Conic Constant	                & -0.227 \\
Vertex Radius of Curvature      & 112.5 mm \\
Aspheric coefficients           & $A=6.59\times10^{-5}$ \\
		                & $B=-5.03\times10^{-7}$ \\
                                & $C=1.58\times10^{-9}$  \\
\hline 
\multicolumn{2}{l}{$^{a}$ Surface parameters are given in terms of}  \\
\multicolumn{2}{l}{$  z =  {c r^{2} \over \left[1 + \sqrt{1 - (1+k)c^{2}r^{2} }  \right] } + Ar^{4} + Br^{6} + Cr^{8}  $ } \\
\multicolumn{2}{l}{where $z$  is the sag of the surface, $c$ is the curvature at} \\
\multicolumn{2}{l}{the vertex, $k$ is the conic constant, and $r$ is the} \\ 
\multicolumn{2}{l}{perpendicular distance from the symmetry axis. } \\
\hline \hline 
\end{tabular} 
\label{tab:optics_specs}
\end{center}
\end{table}

The primary mirror was an off-axis section of a paraballoid, which was
defined as the intersection of a cone and the paraballoid. The apex of
the cone was at the focus of the parabola, had an opening angle of
52\dega , and was centered on an angle of 38\degs from the optical
axis of the parabola.  The focal length from the center of the mirror
to the primary focus was 134~cm.  The mirror was constructed by
Dornier Satellitensysteme (Germany) from a light-weight graphite-epoxy
honeycomb to facilitate modulation and weighs \sima 11~kg.  The
reflecting surface was made of 5000~\AA\ of sputtered aluminum and a
protective layer of 2000~\AA\ of SiO$_{2}$.  Bock measured the
emissivity of samples of the surface near 150 GHz and found values
between 0.27\% and 0.6\% \citet{bock_emissivity}. Contour measurements
gave an RMS surface accuracy of $8.3~\mathrm{\mu m}$ with fixed focus
position and $8.1~\mathrm{\mu m}$ with a shift of focus of
0.61~mm. The mirror was fabricated from the pattern made for the 3-m
Cologne mm-wave telescope to reduce fabrication costs.

The design of the reimaging system was driven by the fixed optical
parameters of the primary mirror, a need for a cooled aperture stop
(Lyot stop) and a requirement for a diffraction limited field of view
of one square degree at frequencies lower than about 420~GHz.  It was
also preferable to cool the secondary optics to reduce the optical
load on the detectors because emission from the telescope was expected
to be a dominant source of detector noise at 150~GHz and 240~GHz,
given the expected \sima 10~\nvrthz amplifier noise; see
Table~\ref{table:opticalload} and Section~\ref{bolonoisechar}.
Cooling the optics required a compact system.

The secondary optics consisted of ellipsoidal secondary and tertiary
mirrors that had aspheric corrections and respective diameters of
21~cm and 18~cm.  They were diamond turned from solid aluminum which
gave an optical quality surface at visible wavelengths.  The back of
each mirror was lightweighted to reduce the mass and heat capacity.
The mirrors were manufactured and assembled inside a baffled optics
box by Speedring Systems (U.S.A.).  Speedring also used laser
interferometry to verify the alignment and optical performance of the
secondary optics. The optics box was housed within the \maxima\
cryostat and maintained at \sima 4~K by liquid helium.  The interior
baffles of the box were blackened with a \sima 0.5~cm thick layer of
combined Stycast 2850 FT black epoxy, carbon lampblack, and 175-$\mu$m
diameter glass beads, which has been demonstrated to be an effective
far-infrared absorber \citep{BockThesis}.

\begin{figure}[ht]
\includegraphics[angle=270,width=3.0in]{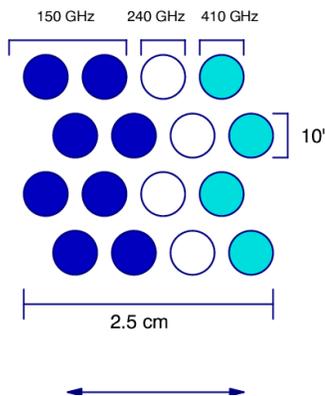}
\caption{The layout of the \maxima\ focal plane as viewed from behind the
bolometers. The arrows indicate the scan direction (azimuth modulation
at constant elevation).  All 16 channels project onto the sky with a
10\mins FWHM beam-size.  The width of the focal plane was 2.5 cm.}
\label{fig:focal}
\end{figure}

We used CODE~V, a software package by Optical Research Associates, to
design and assess the optical performance of the system. The telescope
provided beams with full width at half maximum of 10\mins at all
frequencies; at 150~GHz, 240~GHz, and 420~GHz the worst Strehl ratios
(wave front errors) over the entire focal plane were 0.97 (0.03), 0.94
(0.04), and 0.83 (0.07), respectively. (Strehl ratios (wave front
errors) larger (smaller) than 0.82 (0.06) are considered beyond the
diffraction limit.).  The focal surface was curved for which we
accounted with the placement of the feedhorns. Figure~\ref{fig:focal}
shows a schematic view of the focal plane from the point of view of
the detectors.

We chose a Gregorian system with three foci because it could be well
baffled; see Figure~\ref{fig:optics}. A large Winston cone baffle was
placed outside the cryostat window; a ray arriving from a direction
away from the primary mirror required five edge diffractions with
angles as large as 65\degs to arrive at the focal plane. The Winston
cone baffled was designed to admit throughput only from the primary
mirror.

During observations the primary mirror was modulated around an axis
connecting the center of the primary mirror and its focus point. Less
than 0.4\% of the throughput was modulated on the primary mirror for
any of the beams, minimizing offsets due to temperature or emissivity
gradients.

\begin{table}
\caption{Predicted Optical Load During Flight}
\centerline{\begin{tabular}{cccc}
\hline
 Source & $P_{150}$	& $P_{240}$	& $P_{410}$ \\
	& [pW]	& [pW]	& [pW] \\
\hline
CMB 	&0.11 & 0.10 & 0.01 \\
Primary Mirror$^{a}$ & 0.18 & 0.64 & 0.39 \\
Atmosphere$^{b}$ & $< 0.01$ & $0.03$ & 0.05 \\
Optics Box$^{c}$ & 0.10 & 0.10 & 0.01 \\
\hline Total & 0.39 & 0.87 & 0.46 \\
\hline
\multicolumn{4}{l}{$^{a}$ Assumed primary mirror emmissivities of 0.5\% and}\\
\multicolumn{4}{l}{0.6\% for the 150, 240, and 410-GHz bands, respectively.}\\
\multicolumn{4}{l}{$^{b}$ Atmospheric load was modeled with ATM}\\
\multicolumn{4}{l}{$^{b}$ \citep{pardo}.}\\
\multicolumn{4}{l}{$^{c}$ Approximation (factor of \sima 2).}\\
\end{tabular}}
\label{table:opticalload}
\end{table}


\subsubsection{Lyot Stop}

We located the liquid \4he cooled Lyot (aperture) stop after the
tertiary mirror and before the entrance to the photometers to apodize
the illumination of the primary and to terminate the excess field of
view at \sima 4~K.  The Lyot stop was fabricated from a 0.63-cm thick
sheet of Eccosorb MF-124, a microwave absorber manufactured by Emerson
\& Cuming Microwave Products (U.S.A.).  This thickness provided adequate
absorption of excess radiation \citep{halpern}.

The Lyot stop had an elliptical opening with semi-major and semi-minor
axis lengths of 2.04 and 1.75 cm.  We achieved a well focused beam
despite the thickness of the material by tapering the opening to a
knife edge at an angle of 30 degrees.  The face pointed towards the
sky is covered with a layer of $25 \mathrm{\mu m}$ aluminum foil which
reflected radiation that would otherwise have been transmitted through
the thinnest section of the taper.

\subsubsection{Feedhorns} 

We used a feedhorn coupled array to shield the bolometers from
instrumental optical load and from RFI.  The use of feedhorn coupled
systems is discussed in \citet{griffinhorn}.

The optical signal from the sky was fed to 16 individual photometers
through back-to-back circular aperture copper feedhorns.  At 150~GHz,
10\mins FWHM beams were at the diffraction limit of the telescope.  We
used single moded straight walled feedhorns at 150~GHz.  The
theoretical beam patterns for a single-moded straight walled feedhorn
are well understood \citep{straight}.  The higher frequency channels
were multi-moded and had Winston Cone feedhorns.  The properties of
Winston Cone feedhorns are explained in \citet{Winston}.  All feedhorns
were designed to create 10\mins FWHM beams.  The back-to-back feedhorn
design was used collimate the radiation, which was useful for spectral
filtering.

The opening diameter of the feedhorns was quantified in terms of \NFL ,
where $\lambda$ is the wave length of observation, $F$ is the focal
ratio (f-number) at the focal surface opening, and $N$ is a numerical
factor.  $F$ is determined by the beam FWHM and the reimaging scheme.
The larger $N$ is, the better control one has in defining the beam.
However, increasing $N$ separates the beams centers on the sky, which
increases edge effects on the combined scan region.  We optimized the
aperture efficiency as described in \citet{griffinhorn} by choosing $N
\simeq 2$ (\NFL~=~(6.10:4.47)~mm for the (150:240)~GHz channel
feedhorns).  Though this choice separated the telescope beams by more
than their 10\mins FWHM, our observation strategy did not require the
beams to overlap.

\begin{figure}[ht]
\plotone{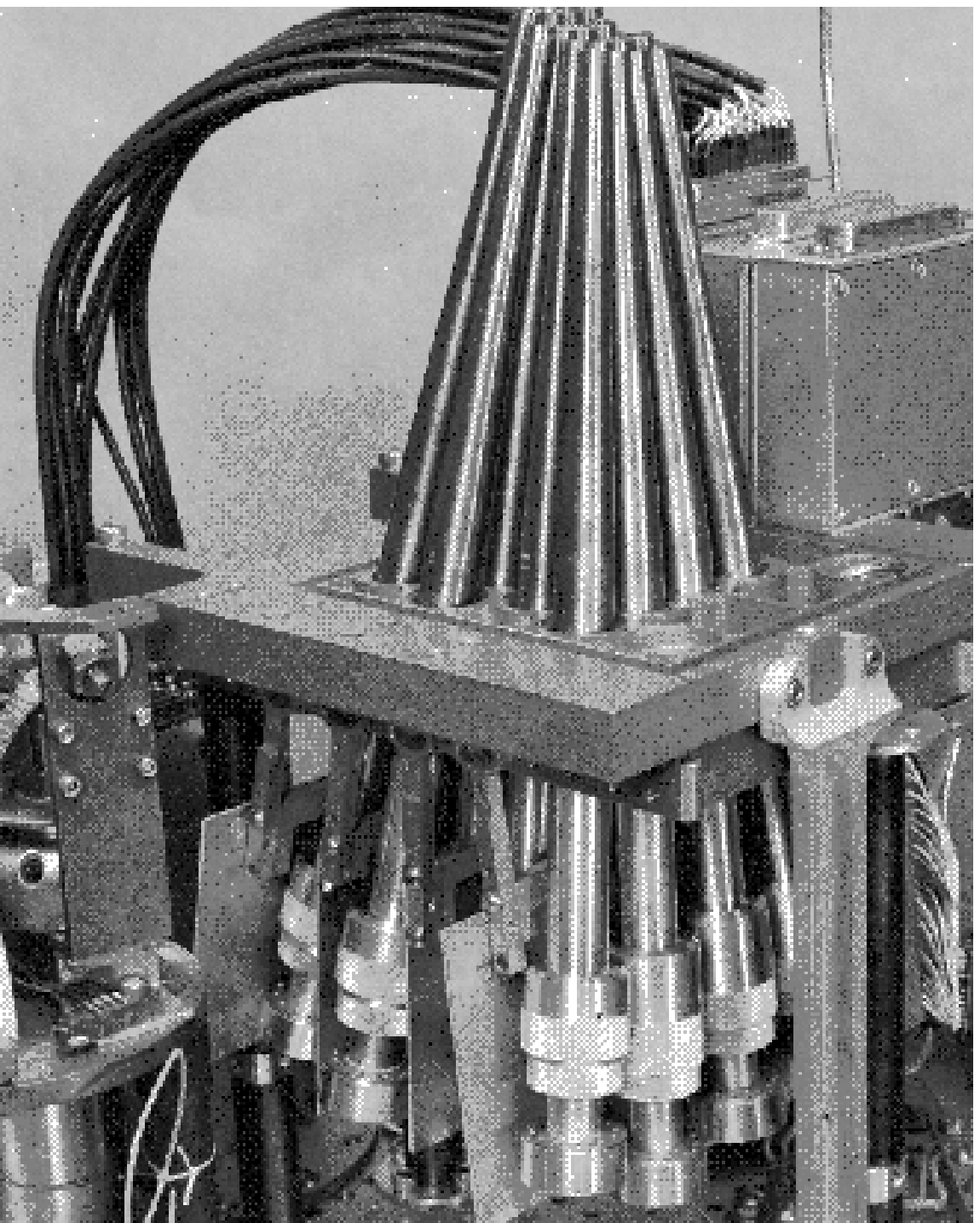}
\caption{Photograph of
feedhorns and bolometer array.  The feedhorns (top of photograph) were
assembled into a gold-plated aluminum plate.  The band-defining metal
mesh filters and bolometers for each photometer were stacked inside
aluminum holders (bottom of photograph).  The holders were assembled
into a gold-plated aluminum plate that supported is by three
thin-walled Vespel SP-22 legs.  The horn and bolometer arrays were
separated by a 0.5 mm gap at room temperature.}
\label{fig:array}
\end{figure}

The feedhorns were fabricated by machining aluminum mandrils to define
the interior surface.  The mandrils were then electroplated with
copper, and the aluminum was etched away in a bath of NaOH.  The
exterior surfaces of the feedhorns were then trimmed on a lathe, and
the interior surfaces were polished.  The feedhorns and bolometer
array are shown in Figure~\ref{fig:array}.

\subsubsection{Frequency bands, filters and detector backshorts} 

\maxima\ observed in frequency bands centered at 150, 240, and 410-GHz.  
The 150 and 240-GHz bands were primarily used to measure the CMB.  The
410-GHz band monitored emission from atmosphere, galactic dust, and
extragalactic infrared point sources.  Data from the three observation
bands were used to discriminate spectrally between various signal
sources.

The frequency bands were primarily defined by metal-mesh filters
located in the relatively collimated beam between the feedhorns and the
bolometers.  The 240 and 410-GHz channels had both high-pass and
low-pass metal-mesh filters.  The 150-GHz channels only had a low-pass
metal-mesh filter; the section of circular wave guide between the
back-to-back straight cones acted as a high-pass filter.

The filters were built from thin metallic meshes supported on Mylar
film substrates by P. Ade of Cardiff University, Cardiff, Wales (then
at QMWC, London, England).  The transmission of the filter was
determined by the pattern and geometry of the film, and the number and
spacing of the stack of meshes.  A thin copper substrate was
evaporated onto a taut $\sim 1.5$ \mum thick Mylar sheet suspended on
a stainless steel ring.  The metal was patterened via photolithograpy
in either a capacitive (low-pass), inductive (high-pass), or resonant
(band-pass) geometry.  The steepness of the cutoff was improved by
stacking multiple meshes separated by $\lambda_0 / 4$, where
$\lambda_0$ is the cutoff frequency of the filter.  These filters can
be repeatedly cryogenically cycled without any stress fractures or
significant performance change.  They have been demonstrated to have
high transmission and sharp cutoffs \citep{LeeThesis}.

The band defining filters leaked slightly at harmonics of the cutoff
frequency.  Three low-pass filters between the cryostat window and the
entrance to the feedhorns blocked these leaks of the band
defining-filters and provided overall rejection at higher
frequencies.  Two of these filters were low-pass metal-mesh filters
with cutoffs at 480~GHz and 570~GHz.  The last was an absorptive
alkali-halide filter, and has a cutoff frequency of 1650~GHz.

The bolometers were suspended in resonant optical cavities with a
depth of $\lambda_{obs} / 4$, created by a flat brass backshort.

\subsubsection{Neutral Density Filter}

We placed a 1\% transmitting neutral density filter (NDF) at the
intermediate focus between the secondary and tertiary mirrors for
optical tests with a 300-K load. The NDF was made by evaporating a thin
continuous metal film onto a taut Mylar sheet stretched over a
stainless steel ring, and was provided by P. Ade.  The NDF was
mounted on an aluminum slider which could be manually moved in and out
of the beam using a vacuum-sealed linear actuator.

\subsection{Pre-flight characterization} \label{pre-flightoptics}

We measured the spectral response and optical efficiency of each
channel in the laboratory before flight.  We measured the beams of the
entire telescope and the far side-lobe response of each channel just
before flight to verify proper focusing and baffling of the
telescope.

\subsubsection{Spectral sensitivity}

The transmission spectrum of each channel was measured before
\maximai\ to verify the performance of the band defining filters.  The
spectra were used to discriminate between foreground and CMB signals
and for calibration from point sources with spectra different from
that of the CMB.

\begin{figure}[ht]
\plotone{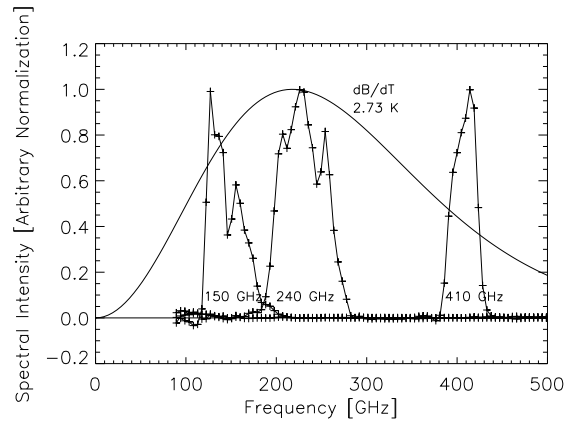}
\caption{The measured, normalized spectral
response for a detector of each color.  The solid curve represents the
derivative of the emission spectrum of a 2.73 K blackbody with respect
to temperature, $dB/dT$.  The spectra were measured before flight with
a Michelson Fourier interferometer.}
\label{fig:m-bands}
\end{figure}

We measured the spectral response from 4.8~GHz to 1.2~THz, with a
resolution of 4.8~GHz using a Michelson Fourier spectrometer
\citep{richardsfts}.  One such spectrum from each of the three
frequency bands is shown in Figure \ref{fig:m-bands}.  The noise in
the spectrum rose significantly below 90~GHz and above 1~THz due to
the reduced efficiency of the beam splitter.  We also saw increased
noise at 540~GHz due to a strong water absorption line.  We estimated a
statistical error of 2\% for the 150-GHz bands, 14\% for the 240-GHz
bands, and 7\% for the 410-GHz bands.

The spectral response for all photometers of a given frequency band
were morphologically similar.  The FWHM band edges averaged over each
set of detectors were 124~GHz to 164~GHz (150~GHz nominal), 199~GHz to
258~GHz (240~GHz nominal), and 393~GHz to 423~GHz (410~GHz nominal).
This gave absolute (fractional) bandwidths of 40~GHz (27.8\%) at
150~GHz, 59~GHz (25.8\%) at 240~GHz, and 30~GHz (7.3\%) at 410~GHz.
The measured band centers varied by several GHz.  In the case of the
150~GHz channels, the measured bands were asymmetric.

We checked for high frequency leaks in the photometers before each
flight by measuring each bolometer's response to a Rayleigh-Jeans
source (a chopped 77-300 K cold load) through a series of
room-temperature high-pass filters.  We placed a conservative estimate
that $\sim 1\%$ or less of the chopped optical power detected by each
bolometer leaked above the high-frequency cutoff of the nominal band
for that photometer.  This was a stringent test of susceptibility to
foregrounds with steeply rising emission spectra over our observation
bands ($B_{\nu} \propto \nu^x, x \geq 2$), such as galactic and
extragalactic dust, the atmosphere, and albedo.

\subsubsection{Optical Efficiency}

The ratio of optical power detected to that which entered the
telescope is less than unity.  This frequency dependent quantity is
the optical efficiency, $\epsilon(\nu)$.  The total optical power
detected, $P_{detected}$, can be expressed as

\begin{equation}
P_{detected} = A \Omega \int_{\nu} \epsilon (\nu) I (\nu) d \nu \simeq A
\Omega < \epsilon (\nu)> \int_{\nu_l}^{\nu_h} I(\nu) d \nu,
\end{equation}

\noindent where $A \Omega = 0.041 \mathrm{cm^2\;sr}$ is the throughput 
of the telescope, and $I(\nu)$ is the spectral intensity of the
observed source.  The average optical efficiency $<\epsilon (\nu)>$ is
defined between FWHM frequencies $\nu_l$ and $\nu_h$.

In the laboratory, we determined the average optical efficiency,
$<\epsilon (\nu)>$, for each channel over its effective bandwidth.
Using the neutral density filter to avoid saturation, we measured the
load curve (\ie\ bolometer resistance versus applied electrical power)
of each bolometer with a 300-K optical load filling the throughput.
We repeated the measurement with a 77-K load.  For points of equal
resistance on the two load curves, the bolometer was electrically
heated to an equal total (electrical plus optical) power; the
difference in electrical power thus gave the difference in detected
optical power from 300-K and 77-K.

We measured only the optical efficiency of the bolometers and the cold
optics, as it was impractical to fill the primary mirror with a
calibrated diffuse source.  The primary mirror had an absorptivity of
order $0.5 \%$ over all observation bands, and should degrade the
overall optical efficiency minimally.

We made this series of measurements twice before \maximai.  Between
measurements, we polished the feedhorns and implemented $\lambda_0/4$
detector backshorts to improve the optical efficiencies.  The
efficiencies achieved are listed in Table \ref{table:o-eff}.

\begin{table}
\caption{Optical Efficiencies for the \maxima\ Receiver}
\centerline{\begin{tabular}{cccccc}
\hline
Channel	& $<\nu>$ & $<\epsilon ( \nu )>$ &Channel	& $<\nu>$ & $<\epsilon
( \nu )>$ \\
	& [Hz]    &  $[\%]$   &	& [Hz]    &  $[\%]$   \\
\hline
b14	&  150   & 18.4& b13	&  240   &3.9  \\
b15	&  150   & 8.4 & b23	&  240   &13.9 \\
b24	&  150   & 20.2& b33	&  240   &9.6  \\
b25	&  150   & 22.8& b43	&  240   &24.2 \\
b34	&  150   & 14.2& b12	&  410   &3.8  \\
b35	&  150   &  -  & b22	&  410   &3.9  \\
b44	&  150   & 16.8& b32	&  410   &3.9  \\
b45	&  150   &  15.0&b42	&  410   &5.6  \\
\hline
\multicolumn{6}{l}{Note.-One channel was not functioning at the time of}\\
\multicolumn{6}{l}{the measurement.}
\end{tabular}}
\label{table:o-eff}
\end{table}

\subsubsection{Focusing}

\begin{figure}[ht]
\plottwo{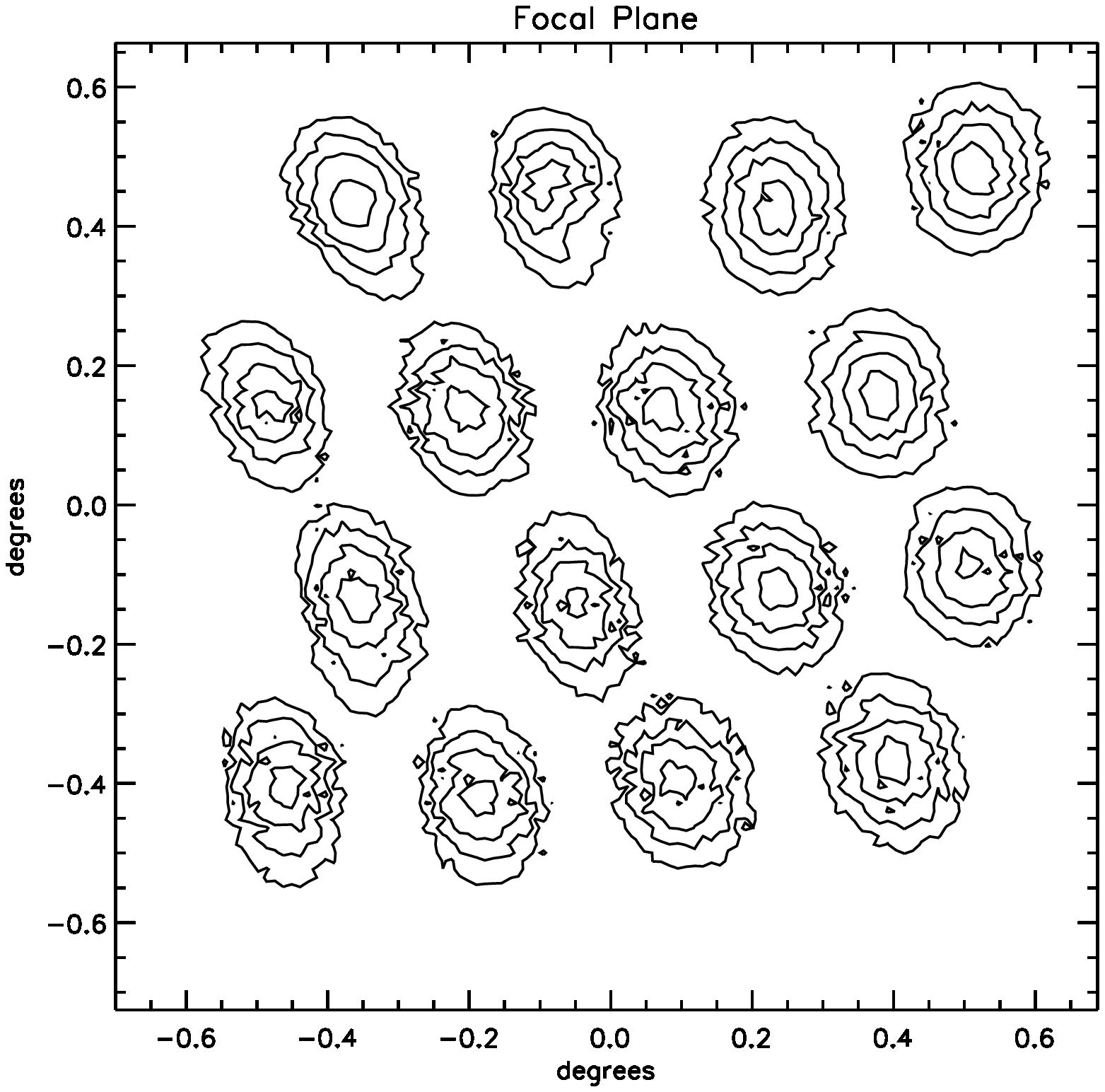}{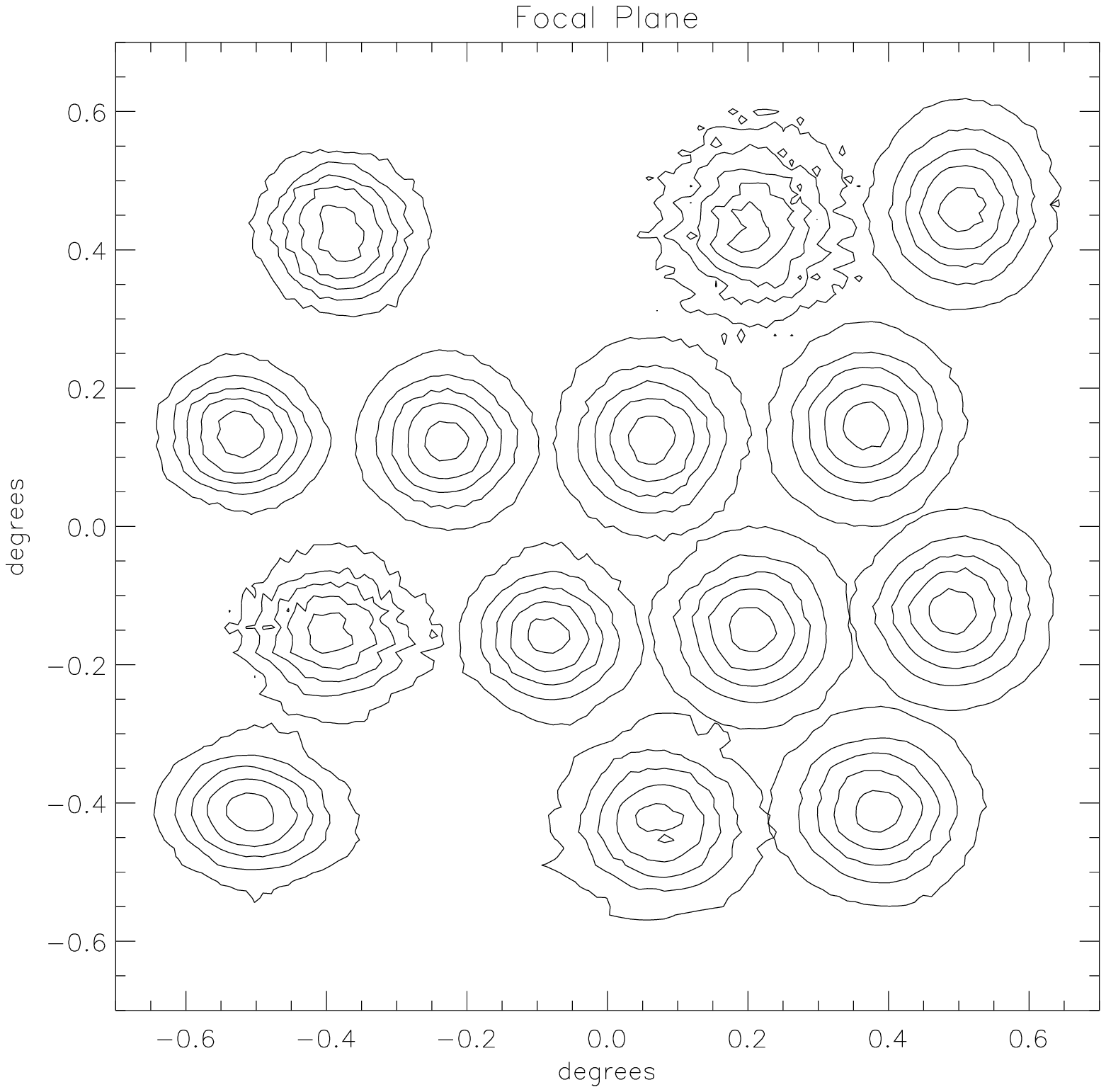}
\caption{Beam contour plots for all channels in
physical units.  The beams are shown as projected onto the sky, which
is the reverse point of view from that shown in Figure
\ref{fig:focal}.  {\bf Left:} \maximai: the contours, moving from each
beam center to each beam edge, represent the $90\%, 70\%, 50\%$, and
$30\%$ levels, respectively.  {\bf Right:} \maximaii: the contours,
from the center of each beam out, represent the $90\%, 70\%, 50\%,
30\%$, and $10\%$ levels respectively.  Two of the 240~GHz channels
were non-operational during the \maximaii\ flight.  Refinements in
focusing techniques between the \maxima\ flights led to improved beam
symmetry in \maximaii.}
\label{fig:allbeams}
\end{figure}

The telescope was focused by bringing the prime foci of the primary
mirror and the secondary optics in the receiver together along a
unique optical axis.  This was initially done using mechanic alignment
tools and visible laser tests.  We then measured the two-dimensional
beam profiles of each channel in the array by performing a raster scan
with the telescope over a $7'$ FWHM source.  The source was a halogen
lamp mounted at the focus of a 1 m diameter $f / 1$ paraboloidal
mirror.  The source was chopped at 5 Hz so that signal appeared above
the 1/f knee in the bolometer signal bands.  Measured profiles were
compared to those predicted from simulations of the optical system
with various types of defocus.  We corrected the focus the telescope
iteratively to minimize beam size and asymmetry.  Beam contours for
both observations are show in Figure~\ref{fig:allbeams}.

\subsubsection{Far side-lobe measurements} \label{sidelobe}

We measured the far side-lobe response of the telescope before flight to
verify that sources such as the Earth, the Moon, the balloon, and the
gondola would not contaminate our measurements of CMB temperature
fluctuations.

We aimed a 150-GHz 20-mW Gunn oscillator at the assembled and baffled
payload from a distance of \sima 20~m.  The oscillator was chopped at
7~Hz.  We performed the measurement outdoors to minimize secondary
reflections off building walls.  Nevertheless, it is likely that we
detected some reflected signals, so the actual side-lobe sensitivity
may be lower than the measurements indicate.

For \maximai, the source was mounted on the top of a 35 m building for
side-lobe measurements as a function of elevation angle.  For
\maximaii, the source was mounted at the end of the arm of a ``cherry
picker'' truck.  The \maximaii\ method more effectively minimized
secondary reflections.  We scanned across elevation and azimuth from the
beam center yielding two orthogonal, one-dimensional side-lobe maps.

The response of the bolometers was linear for signal changes less than
20~dB.  We attenuated the output of the source with both a
dial-attenuator and 1-in (2.54-cm) thick sheets of plywood to keep the
response linear over the 80~dB range of the measurement.  We corrected
for the attenuation of the source by measuring bolometer voltages
before and after each change in attenuation at a fixed angle.

\begin{figure}[ht]
\plotone{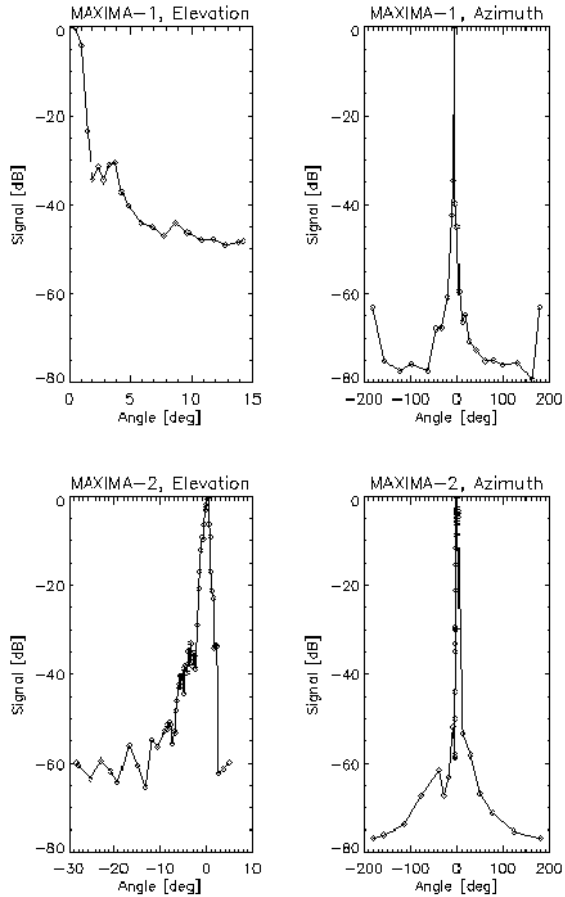}
\caption{Data from pre-flight side-lobe tests.  The source
was roughly 30~m from the telescope.  {\bf{Left Top:}} Test data in
the elevation direction for \maximai.  The angle is that of the
telescope above the test source.
{\bf{Right Top:}} Test data in the azimuth direction for \maximai.
The telescope beam is at fixed elevation (\sima 30\dega).  The source
is moved around the telescope at the same elevation.
{\bf{Left Bottom:}} As above, for \maximaii.  Most of these data were
collected with the source at higher elevation than the telescope beam
(negative angles on the x-axis).
{\bf{Right Bottom:}} As above, for \maximaii.}
\label{fig:sidelobe}
\end{figure}

The measured far side-lobe response is shown in
Figure~\ref{fig:sidelobe}.  The measurement noise floor is roughly
-75~dB.  The elevation beam maps have a higher apparent noise floor,
probably due to reflected signals.  The measured attenuation at
15\degs below the beam is sufficient to prevent significant
Earth-based side-lobe contamination.  The apparent -60~dB back-lobe in
\maximai\ is suspected to be an artifact of the measurement technique.
Even if the back-lobe were real, no sources in the nighttime sky would
be detectable with this attenuation.

\subsection{In-flight characterization of beam patterns} \label{optbeams}

The beam function (\ie the Legendre polynomial expansion of the
beam) is essential for probing CMB power at higher multipoles.  We
measured the beam patterns of each channel in the array during flight
to determine the beam function, \bl.  Beam measurements were also
necessary for point source responsivity calibration.

\subsubsection{Measurement} \label{planet} 

The beam patterns were measured in flight by observing Jupiter for
\maximai\ and Mars for \maximaii.  These planets were both small in
angular size compared to the beam FWHM, and could thus be treated as
point sources.  They were detected with a signal-to-noise ratio of
\sima 100 for Mars and \sima 1000 for Jupiter.

		During planet observations, the telescope tracked the
planet in azimuth while remaining at fixed elevation.  As the planet
drifted through the elevation of the observation, the modulation of the
primary mirror scanned the beams across the planet many times.  As
the planet drifted in elevation the spatial response of each beam was
measured in two dimensions.  The angular offset between the center of
the beam and the boresight CCD camera was also measured in this scan.
The detector data were averaged in fixed azimuth and elevation bins
over a prescribed area and resolution through three-point linear
interpolation.  This beam map was decomposed into the beam function \bl.
We measured the beam FWHM from the $50 \%$ contour of the beam map.

We measured the beams for all of the functioning channels during both
\maxima\ observations.  Two-dimensional beam contours are shown in
Figure~\ref{fig:allbeams}.  The FWHM of the major and minor axes of
each beam is listed in Table~\ref{table:beams}.

\begin{table}
\caption{MAXIMA Beams FWHM}
\centerline{\begin{tabular}{cccccc}
\hline
Channel &$<\nu>$& M1:$\theta_{maj}$ &M1:$\theta_{min}$ &M2:$\theta_{maj}$ &M2:$\theta_{min}$ \\
	& [GHz] & [arcmin]& [arcmin]& [arcmin]& [arcmin]\\
\hline
b14	& 150 &	11.3	&9.0	&10.0	&9.3\\
b15	& 150 &	10.8	&9.0	&10.1	&9.1\\
b24	& 150 &	10.8	&8.4	&9.5	&9.2\\
b25	& 150 &	10.8 	&8.4 	&9.7	&9.3	\\
b34	& 150 &	10.8	&8.4  	&9.5	&9.5	\\
b35	& 150 &	10.0	&8.4	&9.8	&9.4	\\
b44	& 150 &	10.5	&9.2	&10.1	&9.3	\\
b45	& 150 &	10.8	&9.0  	&10.0	&9.4	\\
\hline
b13	& 240 &	10.2	&8.4	&-$^{a}$ 	&-$^{a}$	\\
b23	& 240 &	11.0	&8.4	&8.7	&8.4	\\
b33	& 240 &	11.3	&7.7 	&8.7	&8.2	\\
b43	& 240 &	11.5	&8.4	&-$^{a}$	&-$^{a}$	\\
\hline
b12	& 410 &12.0	&8.3	&8.8	&8.1	\\
b22	& 410 &10.2	&9.0 	&8.3	&7.5	\\
b32	& 410 &	13.6	&7.9	&8.5	&8.1	\\
b42	& 410 &	11.0	&7.7	&9.5	&7.4	\\
\hline
\multicolumn{6}{l}{Note.-$\theta_{maj}$ and $\theta_{min}$ represent the}\\
\multicolumn{6}{l}{major and minor axes.  Values were determined from the}\\ 
\multicolumn{6}{l}{50\% contour for each beam.}\\
\multicolumn{6}{l}{$^{a}$ Non-operational channel}
\end{tabular}}
\label{table:beams}
\end{table}

\subsubsection{Symmetry}

During flight, the telescope revisited regions of the sky at different
times so that the angle of the scan varies due to sky rotation.  For
fully symmetric beams, the same region of the sky was observed upon
revisitation, regardless of angular orientation.  Real beams have some
degree of rotational asymmetry.  Strict correction for this asymmetry
would require that the beam function be expanded in a series of
spherical harmonics (\eg, \citet{Abramovitz}).

Comparing the two panels in Figure~\ref{fig:allbeams}, the beams in
\maximaii\ were noticeably more symmetric than those in \maximai.  Our
approach to this problem was to compute an effective, axially
symmetric beam function, $B_{\ell}$, and to estimate precisely an
error incurred to such a procedure.  \citet{wu_beam} demonstrate that
this approach is accurate for both \maxima\ flights.

\subsubsection{Beam Characterization Uncertainty}

We determined that the beam size uncertainty causes less than $4 \%$
and $11 \%$ uncertainty in the \cl\ estimates for $\ell$ = 410 and
785, respectively, and that beam asymmetry does not contribute
significantly to power spectrum errors at any angular range.  Complete
beam errors contributions are reported in \citet{lee_results} and
\citet{hanany_results}.

Beam map uncertainties resulted from errors in the pointing
reconstruction, detector noise, the uncertainties in the bolometer
electronic filters and time response, and the limited area and
resolution of the beam map.  We list the contributions to the total
error from these sources to the measured beam FWHM for one 150~GHz
channel in Table~\ref{table:error}.  The total error is the quadrature
sum of the individual contribution.

\begin{table}
\caption{Sources of Beam Error for one 150 GHz Channel}
\centerline{\begin{tabular}{lc}
\hline
Source of Error	&                $\sigma_{fwhm}$ \\
	& [arcmin] \\
\hline
White Noise in the Detector Time Stream &  0.09\\
Bolometer Electronic Filters &       0.13  \\
Bolometer Time Constant           &  0.06\\
Phase Preserving Low-pass Filter         &        0.14 \\
Pointing Reconstruction (differential err.)     &  $ 0.05$\\
Resolution of Beam Map &         0.17 \\
Area of Beam Map        &        0.15 \\
\hline
Total			& 0.32 \\
\hline
\end{tabular}}
\label{table:error}
\end{table}

\section{DETECTORS}\label{detectors}

The \maxima\ detector system was a 16-element bolometer array.  A
bolometer is a thermal detector whose electrical resistance varies as
a function of temperature.  This resistance is read out electrically.
Thorough reviews of bolometers can be found in \citet{bolo_review} and
\citet{galeazzi}.

The audio frequency bandwidth of the bolometer signal must be wide
enough to avoid constraining the $\ell$-space coverage in the
measurement of the power spectrum.  The signal bandwidth is determined
by the 1/f knee and the high frequency roll-off in the response of the
instrument.  Though some of the cosmological signal appeared at
frequencies as low as the telescope scan frequency of \sima 20~mHz, it
was primarily found above the 0.45~Hz frequency of the primary mirror
modulation.  Cosmological signals appeared as high as the the roll-off
frequency of beam window function, as determined by the telescope scan
speed.  We set the high frequency roll-off of the electronic filters
at 20~Hz, above the beam roll-off.

Low-frequency 1/f noise may cause striping in the maps, but it was
effectively suppressed by cross-linked observations; sensitivity to
large scale CMB fluctuations was mostly limited by observation area
(sampling variance) but was also affected by residual 1/f noise.

The bolometer response time constrains the high-frequency cut-off of
the signal band and limits the choice of scan speeds for the
telescope.  A slow detector response compared to the telescope scan
speed has the effect of smearing the beams on the sky, distorting the
window function.  \citet{hanany_tc} shows that to prevent significant
distortion, bolometer response time must be at least 2.5 times faster
than the time to scan a single beam size.  This corresponded to a
detector response time of 10~ms or less for our scans and beams.

\subsection{MAXIMA bolometers}

\begin{figure}
\plotone{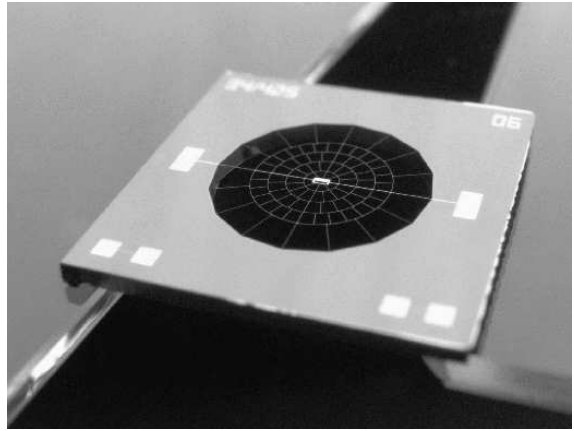}
\caption{The bolometer pictured is of the
same design as those used in \maxima.  The spider-web absorber has a
2.5 mm diameter with a 5\% filling factor.  The radial components
within the spider-web pattern are 160~\mum long and 4 \mum wide.  The
NTD-19 thermistor, a cube with 250~\mum sides, is bump bonded to the
center of the web.  The chip is powered through two lithographed gold
leads.  Mechanical support comes from the electrical leads and sixteen
\sin\ legs, 1~mm long and 5~\mum wide.  Photograph courtesy of
J. Bock.}
\label{fig:maximabolo}
\end{figure}

We used composite bolometers obtained from J. Bock at JPL
(Figure~\ref{fig:maximabolo}).  The bolometers were made with metalized
mesh absorbers and neutron transmutation doped germanium (NTD-Ge)
thermistors, the latter of which are obtained from J. Beeman and
E. Haller at LBNL, Berkeley.

The temperature dependence of the electrical resistance of doped
semiconductor thermistors has the form, $R(T) = R_0 \exp{(\Delta /
T)^{\frac{1}{2}}}$, where $\Delta$\ \sims 10~K and is determined by the
type and level of doping in the material.

The characteristic temperatures for the various types of NTD-Ge allow
for operation between temperatures of 20~mK to 4.2~K.  \maxima\
bolometers were made from NTD-19, which is optimized for operation at
\sima 100~mK.  More information on neutron transmutation doping can be found
in \citet{ntd}.

Electrical contacts on the NTD-Ge chips were made through ion
implantation, followed by a thermal annealing process.  The
wafers were metalized with a $\sim 20$ nm-thick layer of
Cr or Ti, followed by a $\sim 150$ nm layer of Au for easy attachment
of wires.  The wafers were then diced, etched, and passivated.  Gold
leads were indium bump-bonded onto the metalized edges of the
thermistor.

The bolometer absorbing structure was fabricated via optical
lithography starting with a silicon on insulator (SOI) wafer coated
with a 1 \mum layer of silicon nitride (\sin).  The top of the \sin\
was coated with a layer of gold whose thickness was selected to obtain
an average sheet resistance of $377 \Omega / \Box $.  A web with this
sheet resistance absorbs 50\% of the incoming radiation in one pass
\citep{bock_bolo}.  The absorber was then patterned in a `spider-web'
geometry with photoresist.  A dry-etch was used to remove the gold and
around the pattern.  The SOI was dry etched from the back side of the
wafer with deep trench Si etcher.

The spacing of the absorber legs was small enough to trap
millimeter-wave radiation, but was large enough to let high energy
cosmic rays pass through.  The absorber had a 5\% filling factor,
which reduced its cosmic ray cross-section by \sima 90\%.  The optical
efficiency was further increased by placing the bolometers in an
integrating cavity with a characteristic depth of $\lambda / 4$.

\subsection{Noise Characterization}\label{bolonoisechar}

The major contributors to bolometer noise are photon noise, Johnson
noise, thermal fluctuation noise, and noise from the first stage of
amplification (amplifier noise).  A thorough discussion and derivation
of each noise term is found in \citet{bolo_review}.  Bolometer noise is
generally discussed in the context of noise equivalent power, NEP, (or
noise equivalent temperature, NET) which is defined as the incident
signal power (or temperature) required to generate a signal equal to
the noise in a one Hz bandwidth.  The individual noise sources are
uncorrelated.  The total noise is given by the sum of each term in
quadrature.

\begin{figure}[ht]
\plotone{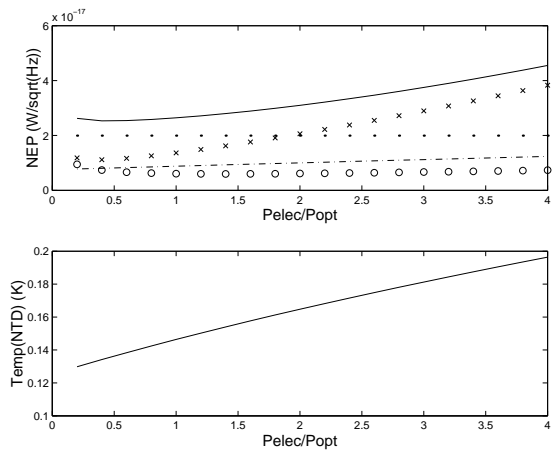}
\caption{Bolometer noise simulation.  The
top panel is a plot of the contributions to the bolometer NEP as a
function of the ratio of the electrical power, $P_{elec}$ to the
optical power, $P_{opt}$.  The contributions are thermal fluctuation
noise (dot-dash), Johnson noise (circles), and amplifier noise
(crosses), photon noise (dotted).  The solid line is the quadrature
sum of the terms.  The bottom panel is a plot of the thermistor
temperature as a function of the same ratio of powers.  We assume an
operating temperature, $T_{base}$ of \sima 100~mK, an optical load,
$P_{opt}$, of 2~pW, an average thermal conductance, $G$ of 70 pW~K$^{-1}$, and
JFET voltage noise, $e_n$ of 6\nvrthz.}
\label{fig:bolonoiselow}
\end{figure}

We choose the thermal conductance $G$ required to reduce the
contribution of thermal fluctuation noise compared to photon noise
(Table~\ref{table:bolog}).  We operated the bolometers at a bath
temperature of $T_0$ \approxs 100~mK, which is cold enough that thermal
fluctuation noise is less than the photon noise.  We used low noise
JFET pre-amplifiers.  We constrained these various parameters with the
help of computer-generated models, such as the one shown in
Figure~\ref{fig:bolonoiselow}.

\begin{table}
\caption{Average Bolometer Thermal Properties}
\centerline{
\begin{tabular}{ccc} \hline 
$Observation$ & $Thermal$ & $Temperature$ \\ 
$Frequency <\nu>$ & $Conductance <G>$ & $Rise <\frac{\Delta T}{T_0}>$ \\ 
$\mathrm{[Hz]}$ & [pW~K$^{-1}$]& \\
\hline
150	&71	& 0.51 \\
240	&290	& 0.29 \\
410	&320	& 0.21 \\
\hline
\end{tabular}}
\label{table:bolog}
\end{table}

\begin{figure}[ht]
\plotone{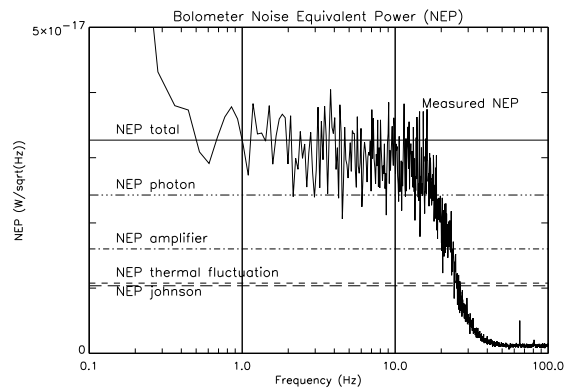}
\caption{The measured noise equivalent power
(NEP) spectrum for a 150~GHz \maxima\ bolometer plotted along with
theoretical NEP contributions from photon (dotted), Johnson (long
dash), thermal fluctuation (short dash), and amplifier (dot-dash)
noise.  The total NEP (solid line) is the quadrature sum of the terms.
Low frequency noise below 0.4 Hz arises from temperature drifts in the
cryogenic system.  Electronic filters in the readout circuit cause the
signal to roll off above 20 Hz }
\label{fig:bolonoise}
\end{figure}

We computed bolometer noise power spectra from a subset of the data
during the CMB observations.  We found the frequency power spectrum
out to the Nyquist frequency of 104 Hz.  A typical noise power
spectrum is shown in Figure~\ref{fig:bolonoise}.  The spectrum can be
divided into three frequency ranges.  For frequencies below \sima
100~mHz, the spectrum decreases as 1/f.  At intermediate frequencies,
the noise spectrum is white.  At frequencies above $\sim 20$ Hz, the
spectrum decreases because of the electronic low-pass filter and
bolometer time constant.  Noise measurements for all functioning
channels are presented for \maximai\ in Table~\ref{table:bolonep1} and
for \maximaii\ in Table~\ref{table:bolonep2}.  The voltage noise was
determined from the average noise measured around a narrow audio
frequency band ($7.0 \pm 0.2$ Hz) near the center of the bolometer
signal band.  We assigned an uncertainty of \sima 2\nvrthz to conservatively account for the non-uniformity of the
white noise level over the signal band.  This uncertainty did not take
into account isolated peaks in the noise spectra of microphonically
sensitive detectors (see discussion below).  These measurements served
as a diagnostic of in-flight instumental performance.  A more
sophisticated noise estimation, as described in
\citet{stompor_mapmaking}, was used for finding CMB maps and angular
power spectra.  Tables~\ref{table:bolonep1} and~\ref{table:bolonep2}
include the overall NEP and NET for the channels grouped by
observation frequency.

\begin{table}
\caption{MAXIMA-I Bolometer Characterization}
\centerline{\begin{tabular}{cccccc}
\hline
Channel &$\nu_{obs}$&$V_n$		& $NEP$  & $NET_{cmb}$ &$\tau$ \\
	& [GHz] &[\nvrthz] & $[\times 10^{-17}$~\Wrthz]
	&$\mathrm{[\mu K \sqrt{sec}]}$&[msec] \\
\hline
b14	& 150 & 16  	&2.8 	&130  	&12.5	\\
b15	& 150 & 10  	&2.0 	&80  	&12.5	\\
b24	& 150 & 18  	&4.0 	&170  	&12.5	\\
{\bf b25}& 150 & 12  	&2.4 	&80     &10.0	\\
{\bf b34}& 150 &12  	&1.8 	&75     	&7.0	\\
b35	& 150 &22  	&4.6 	&200  	&9.5	\\
b44	& 150 &38  	&6.8 	&230  	&10.5	\\
{\bf b45}& 150 &12  	&2.2 	&80     	&9.0	\\
$150_{comb}$	& 150	&5  	&0.9 	&35  	&-	\\
\hline
b13	& 240 &14  	&3.4 	&380      	&10.5	\\
b23	& 240 &14  	&6.0  	&185  	&7.0	\\
{\bf b33}& 240 &10  	&5.4  	&140     	&9.5	\\
b43	& 240 &24  	&9.8 	&345  	&9.0	\\
$240_{comb}$	& 240	&7  	&2.4 	&100  	&-	\\
\hline
b12	& 410 &18  	&6.6 	&-	&12.0	\\
b22	& 410 &20  	&11.0  	&-	&8.5	\\
b32	& 410 &24  	&7.6	&-	&6.0	\\
{\bf b42}& 410 &34  	&19.0  	&-	&5.0	\\
$410_{comb}$	& 410	&11  	&4.4  	&-	&-	\\
\hline
\multicolumn{6}{l}{Note.-Data from the channels in {\bf bold} were}\\
\multicolumn{6}{l}{used for the final data analysis.}
\end{tabular}}
\label{table:bolonep1} 
\end{table}

\begin{table}
\caption{MAXIMA-II Bolometer Characterization}
\centerline{\begin{tabular}{cccccc}
\hline
Channel &$\nu_{obs}$&$V_n$		& $NEP$  & $NET_{cmb}$ &$\tau$ \\
	& [GHz] &[\nvrthz] & [$\times 10^{-17}$~\Wrthz]
	&$\mathrm{[\mu K \sqrt{sec}]}$&[msec] \\
\hline
b14	& 150 &28      	&7.6  	&345  	&12.5	\\
b15	& 150 &10   	&2.4  	&65  	&11.5	\\
b24	& 150 &10   	&3.0 	&100 	&10.5	\\
{\bf b25}	& 150 &10  	&2.8  	&80  	&12.0	\\
{\bf b34}	& 150 &10  	&2.6 	&100  	&6.5	\\
{\bf b35}	& 150 &8  	&2.4  	&90  	&8.0	\\
b44	& 150 &28  	&8.8 	&350  	&8.5	\\
{\bf b45}	& 150 &10  	&2.8 	&85  	&8.5	\\
$150_{comb}$	& 150	&4   	&1.0 	&35  	&-	\\
\hline
b13	& 240 &	-	&-	&-	&-	\\
b23	& 240 &	10  	&8.0 	&165  	&5.0	\\
b33	& 240 &	12  	&6.0 	&170  	&8.5	\\
b43	& 240 &	-	&-	&-	&-	\\
$240_{comb}$	& 240	&8  8 	&4.8  	&120  	&-	\\
\hline
b12	& 410 &	20  	&18  	&-	&14.0	\\
b22	& 410 &	12  	&16  	&-	&6.5	\\
b32	& 410 &	12  	&14  	&-	&2.5	\\
b42	& 410 &	20  	&24  	&-	&3.0	\\
$410_{comb}$	& 410	& 7   	& 8.1 	&-	&-	\\
\hline
\multicolumn{6}{l}{Note.-Data from the channels in {\bf bold} were}\\
\multicolumn{6}{l}{used for the \citet{abroe_results} and}\\
\multicolumn{6}{l}{\citet{radek_cras} data analysis.}
\end{tabular}}
\label{table:bolonep2}
\end{table}

As the overall NEP for each channel was largely determined by a subset
of the best performing bolometers, only these few were used for the
published \maxima\ data analysis.  Maps of CMB temperature anisotropy
were made using three 150~GHz channels and one 240~GHz channel for
\maximai\ and from four 150~GHz channels for \maximaii.  Data from
these channels and from one 410~GHz channel were used for systematic
tests, notably spectral discrimination of foreground signals.

For most \maximai\ channels, noise during CMB observations was
stationary at the 10-20\% level.  A subset of the channels had high
microphonic sensitivity and for these the NET value, which was
calculated using a section of the data without microphonic noise, does
not represent the effort involved in, or the expected benefit of
analyzing the entire data set.  These channels were not included in
the final analysis.  Most timestreams exhibited some degree of noise
correlated with the spatial modulation of the primary mirror.  This
mirror synchronous signal proved to be stationary over long enough
time scales, and we developed a formal template subtraction in
subsequent analysis \citep{stompor_mapmaking}.  Low frequency noise
rose as 1/f, with knees between 0.2~Hz and 0.6~Hz.  Detector noise was
Gaussian except near the chopper frequency.  Frequencies below 0.1~Hz
were poorly sampled and have been marginalized over in data analysis.

For \maximaii, low frequencies noise rose more steeply than 1/f in all
channels, with knees ranging from 1~Hz to 3~Hz.  The source of the
noise is unknown, and the time scale over which it was stationary
varied between channels.  In addition, mirror synchronous noise
appeared in the higher observation frequency channels, especially
during the beginning of the second CMB observation.  The reported
$NEP$s and $NET$s are not a complete reflection of the overall noise
performance.  Analysis of \maximaii\ data (\eg\ \citet{abroe_crosscor})
required marginalization over frequencies below 0.2~Hz.

\subsection{Response Time Characterization}

Measurements of the bolometer response times and noise were made
during both the \maximai\ and \maximaii\ flights.  We determined the
response times by deconvolving a single pole low-pass filter, whose
characteristic frequency is a free parameter, from bolometer time
streams while scanning the primary telescope back and forth in azimuth
across a planet.  The bolometer time constant parameter was adjusted
until the left and right going scans spatially coincided.  This
measurement had a statistical error of 0.50~msec.  The measurements
are presented in Tables~\ref{table:bolonep1} and~\ref{table:bolonep2}
for \maximai\ and \maximaii\, respectively.  We replaced detectors in
many channels between the two flights, which explains the difference
in response time for many channels.  In all cases, the measured
response time was small enough to prevent significant beam smearing
from the telescope modulation.

\section{RECEIVER AND ELECTRONICS}\label{receiver}

\begin{figure}[ht]
\plotone{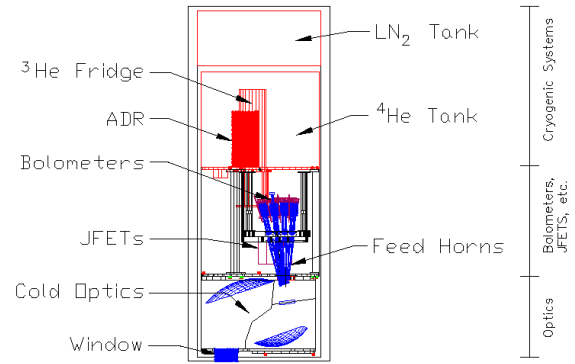}
\caption{A mechanical drawing cutaway of the
\maxima\ cryogenic receiver.  The bottom section of the receiver
contained baffled, liquid \4he-cooled optics and the internal relative
calibration source.  The middle held the bolometers and bolometer
feedhorns, cryogenic JFET preamplifiers, and thermal switches.  The
top section of the receiver contained the cryogenic systems, with the
low temperature refrigerators surrounded by the liquid \4he tank.}
\label{fig:receiver}
\end{figure}

		The \maxima\ cryostat (Figure~\ref{fig:receiver}) housed
the secondary optics, the bolometer array and preamplifiers, an
optical calibration source, and the cryogenic system.  There were also
a number of diagnostic devices including `dark' detector channels not
exposed to the CMB and a variety of internal temperature monitors.

The cryostat was manufactured by Infrared Laboratories (model HDL14).
It had an outer diameter of 40.5~cm and a length of 112.5~cm.
Liquid~\n2 and \4he were held in stainless steel-walled tanks, which
were structurally supported by G-10 stand-offs.  The coldplate for
each tank was made of 1.9~cm thick gold-plated OFHC copper, which
provides excellent thermal contact.  The joint between the copper
coldplate and the stainless steel wall of the \4he tank degraded after
repeated thermal cycling over six years.  We believe the weakness of
the joint was due to differential thermal contraction over the large
size of the tank.  A replacement tank and coldplate were both
constructed from aluminum to minimize differential thermal
contraction.

\subsection{Cryogenics}\label{receiv:cryo}

The detector array was cooled to \sima 100~mK during flight via a four-stage
refrigeration process.  The array was housed inside an evacuated
cryostat, and was cooled by an adiabatic demagnetization refrigerator
(ADR), which dumped its heat to a \3he sorption refrigerator backed by
a pumped liquid \4he bath, and a pumped liquid \n2 bath.  These
cooling systems are summarized in Table~\ref{tab:holdtimes}.

\begin{table}
\caption{\maxima\ Cryogenic Systems}
\begin{center}
\begin{tabular}{ccccc}
\hline 
Cooler & Liquid~\n2 & \4he & \3he & ADR 
\\\hline
Temperature (K) & 50 & 2-3 & 0.35 & 0.1
\\Hold Time (Hours) & 24 & $>$30 & $>$36 & 12
\\Thermal Cycle & Open & Open & Closed & Closed
\\ \hline
\multicolumn{5}{l}{Note.-Numbers are quoted for flight conditions.}
\end{tabular}
\label{tab:holdtimes}
\end{center}
\end{table}

	A 13-liter Liquid~\n2 tank cooled an outer layer of radiation
shielding to 77~K.  This temperature dropped to \sima 50~K when the
Liquid~\n2 tank was exposed to vacuum, as in flight.  The Liquid~\n2
temperature radiation shields were covered with thin, low emissivity
aluminum foil.

	Inside the Liquid~\n2-cooled space was a 21-liter liquid \4he
tank and an additional layer of shielding at liquid \4he temperature.
The outer shell of the cold optics box served as part of this
radiation shielding.  The outside of these shields was low emissivity
aluminum, while the inner surfaces were coated with a blackening
mixture.  The blackened interior absorbed high temperature radiation
that leaked past the shields.

	Within the liquid \4he temperature space were the optics, the
detectors, the JFET preamplifiers, the sub-Kelvin coolers, and a
variety of thermometers.  The optics and most electrical components
were thermally linked to the coldplate.

	Various locations in the liquid \4he space ranged in
temperature from 4~K to 6~K, depending on thermal load and proximity
to the helium tank.  When the liquid \4he was exposed to vacuum, for
testing or in flight, these temperatures drop to 2~K to 3~K.  This
caused a significant drop in the background loading of the bolometers.

	Inset into the liquid \4he tank were the adiabatic
demagnetization refrigerator (ADR) and the liquid \3he refrigerator.
All wiring entering the receiver passed through the liquid \4he and
Liquid~\n2 tanks and was made of low thermal conductivity stainless steel
leads.
		
\subsubsection{\3he Refrigerator}

We designed and constructed a large capacity \3he sorption
refrigerator for the \maxima\ cryostat.  The closed-cycle
refrigerator consisted of 42 STP liters of \3he of $99.995 \%$ purity,
an activated charcoal sorption pump, a copper condenser, and a copper
evaporator which was hermetically coupled to the condenser with a
thin-walled stainless steel bellows.  The bellows was structurally
reinforced with three thin-walled Vespel SP1 tubes connecting the
evaporator to the \4he cold-plate.

Due to size constraints within the cryostat, the sealed \3he
refrigerator was coupled to an external tank to prevent
overpressurization.  The pressure of the combined vessels at room
temperature was 745 KpA; a factor of 4 times smaller than the elastic
limit of the bellows.  The condenser was directly coupled to the \4he
coldplate.  The sorption pump consisted of 131~gm of activated
charcoal.  The grains of charcoal were glued with Stycast to closely
spaced copper fins which maximize the pump area.  Calculations show
that optimal sorption is achieved with at least 3 grams of charcoal
per STP liter of gas (Duband (1990), private communication).

The pump was thermally linked to the \4he coldplate with a ultra-high
purity tin wire.  The thermal conductivity of tin decreases
rapidly with temperature.  The tin wire was a poor thermal link for
temperatures above 10~K.  The refrigerator was cycled by
first heating the charcoal to 40~K for \sima 30~minutes, which
desorbed the \3he from the pump.  After heating, the charcoal and tin
slowly cooled back to the \4he coldplate temperature.  Meanwhile, the
desorbed atoms condensed into liquid and pooled into the evaporator.
Once the charcoal had cooled below 10~K, it pumped on the
\3he, which cooled the liquid to \sima 300~mK.  The refrigerator
maintained this base temperature for 1.5-2 days with an unpumped \4he
bath, and had a cooling capacity of 25~J.

\subsubsection{ADR}

We used of an adiabatic demagnetization refrigerator (ADR) developed
in Berkeley as a prototype for the Space Infrared Telescope Facility
(SIRTF).  This ADR was flown multiple times on the \max\ experiment.
The details of the construction and testing of the
\max /\maxima\ ADR are found in \citet{timbie} and \citet{saltpill}.

The ADR consisted of a paramagnetic salt pill inside a superconducting
magnet.  The salt pill was made from 40~gm of ferric ammonium alum
(FAA), and was supported within the shielded coil with a kevlar string
suspension that was thermally intercepted by the \3he refrigerator
cold-stage.  The \4he cooled superconducting Nb-Ti coil generated a
peak field of 2.5~T with a current of 6.2~A.  The ADR was thermally
cycled with a remotely commandable current controller.  During
cycling, the heat of magnetization was dumped to the \3he refrigerator
cold-stage.  The ADR had a laboratory tested hold-time of 17~hours and
a flight-tested hold-time of 12~hours at \sima 100~mK with a $\sim 90 \%$
duty cycle.  The ADR had a cooling capacity of 0.093~J.

The wiring from the outside of the cryostat to the magnet was designed
to minimize thermal conductivity and electrical power dissipation.
The magnet current entered the cryostat through hermetic connectors
and was carried by bundles of standard 22 AWG copper wire heat-sunk to
the liquid \n2 coldplate.  Each bundle of wire was soldered to a
high-Tc superconducting lead (YBCO) (Eurus Monoco (U.S.A.)) clamped
between the liquid \n2 cold-plate and the top of the liquid \4he tank.
The magnet current was then carried by formvar coated copper clad
Nb-Ti superconducting wire which was heat-sunk to the side of the
liquid \4he tank.  The Nb-Ti leads were potted in Eccosorb CR-124
before entering the \4he cold-plate area, in order to filter RFI.

All of the wiring from the high-Tc leads to the magnet coil was
superconducting during cryogenic operation.  Any significant
electrical power dissipation was dumped into the liquid \n2.

\subsection{Bolometer wiring}

The signals for each bolometer entered the cryostat via hermetic
connectors.  Once inside the cryostat, the signals were carried by a
harness of low thermal conducting stainless steel wire (Cooner Wire
(U.S.A.)) heat-sunk at its midpoint to the liquid \n2 thermal stage.
The signals were then carried in Kapton wire striplines of 50~\micro
gold plated Ni wire, fabricated by Tayco Co. (U.S.A.), which were
potted in Eccosorb CR-124 (Emerson \& Cuming Microwave Products
(U.S.A.)).  These modules were heat-sunk to the
\4he cold-plate and blocked radio frequency interference signals
(\S\ref{receiv:rfi}).

The wiring between the RFI filters and the JFET modules
(\S\ref{readout}) was standard Teflon-coated multi-strand copper.  The
signals were then carried from the JFET module to the \sima 100~mK
array via formvar coated 50~\micro platinum tungsten wire
(California Fine Wire (U.S.A.)), which had a low thermal conductance.
The wiring between the JFETS and the bolometers was susceptible to
microphonics.  We bundled and twisted the wires for the bolometers
around four thin walled G-10 tubes that were rigidly connected to both
the JFET module and the array.  The tubes were thermally intercepted
by an OFHC copper heat strap from the 0.3 K thermal stage.

The wiring continued from the G-10 tubes to each photometer on four
G-10 printed circuit boards with 0.25 mm copper traces.  We covered both
sides of the circuit board with electrically insulated aluminum tape
to reduce the emissivity of the array.  More details on wiring inside
the \maxima\ cryostat can be found in \citet{CDWThesis}.

\subsection{Housekeeping thermometry}

We monitored the internal temperatures with various cryogenic
thermometers; we used silicon diodes at 77~K, carbon composition
thermistors at 2~K to 4~K, an RuO$_2$ thermometer for the \3he
refrigerator (0.3~K), and a germanium thermistor for the ADR
(0.1~K).  Thermometer leads entered the cryostat through the same
hermetic connectors as the bolometer signals.  They were then carried
by the same stainless steel wire harness and Eccosorb RFI filter
module into the cold-plate area.  Each device was wired to the RFI
module with low thermal conducting Teflon coated 0.125 mm manganin wire
(California Fine Wire (U.S.A.)).  The wires for the 0.1 K thermometer
were heat-sunk midway on the \3he refrigerator cold-stage.

\subsection{Internal Relative Calibrator}\label{receiv:stim}

		Bolometer responsivity varied over the duration of a
flight, primarily because of variations in the temperature of the
\sima 100-mK ADR.  These variations were monitored using a stable
internal calibration source (\ie\ a stimulator) consisting of a thin
nickel-chromium layer (2~mm $\times$ 2~mm) backed with a sapphire
substrate.  The metal layer was impedance matched to radiate
efficiently into free space when heated.  When a heating current \sima
1~mA was applied, the metal warmed to \sima 50~K with a time constant of
\sima 1~sec.  The heating current was maintained for 10 seconds, and was
applied every 20 minutes during flight.  The stimulator was mounted
inside the cold optics box, and was fitted with a light pipe to illuminate
the focal plane array from just outside the optical path.  The
illumination of the array was not uniform, with detectors closer to the
calibrator receiving about twice the flux of the more distant
detectors.

	The source was extremely stable with negligible resistance
fluctuations and $<$1\% current fluctuations.  The on-state
calibrator temperature was further stabilized by a weak, temperature
dependent thermal link to the liquid helium stage.

	The absolute flux was not well measured, so the stimulator was
used purely for monitoring of responsivity variations over time.
Absolute calibration was obtained from celestial sources (the CMB
dipole and planets).  The use of stimulator data in detector
calibration is discussed in \S\ref{calib:relative}.

\subsection{RFI protection}\label{receiv:rfi}

		During flight, several radio transmitters were used for
telemetry.  Each radiated 15 to 40 Watts at frequencies of 1.5~GHz and
higher.  Extensive filtering was required to prevent pickup from these
sources in the detectors.

		The receiver itself consisted of three metallic shells
which served as partial Faraday cages.  External cabling was also
fully enclosed in a metallic shell.  All cables between the receiver
and the readout electronics and all cables exiting the readout
electronics passed through commercial RF filtered connectors (Amphenol
FPT02 Series; 60-dB attenuation at 1.0~GHz).  Within the receiver, all
wiring was potted in 27~cm of Eccosorb CR-124, a metal-filled epoxy
commercially available from Emerson and Cuming, that acts as a radio
frequency low-pass filter (30-dB attenuation at 1.0~GHz).

\begin{figure}[ht]
\plotone{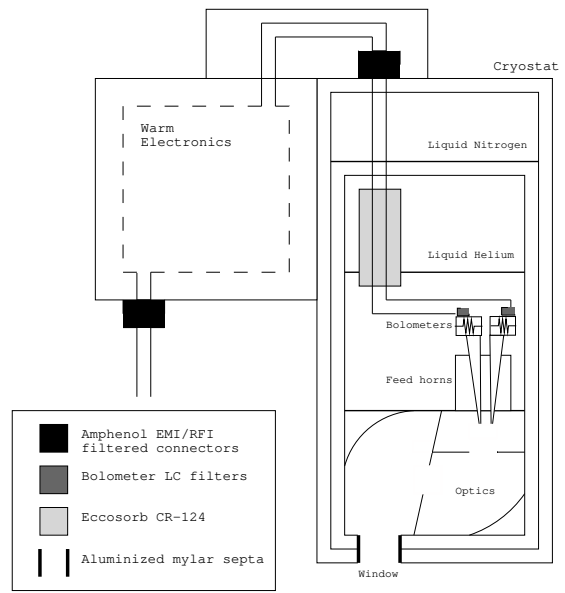}
\caption{The bolometers (shown with resistor symbols)
were shielded from electrical RFI pickup by three different types of
filters in series, shown as filled rectangles.  RFI was prevented from
coaxially transmitting between the temperature baffles by contiguous
layers of aluminized mylar placed around the optical path at the
cryostat window.  The cryostat and the warm electronics shield acted as
Faraday cages.}
\label{fig:allrf}
\end{figure}

The optical window acted as a high-pass filter for ambient RFI, which
had a measured cut-off of \sima 2.5~GHz.  Higher frequency RFI in the
optical path was blocked by taping a cylindrical sleeve made from
multiple layers of 25~\micro sheets of aluminized mylar
around the perimeter of the optical path near the cryostat window with
aluminum tape (Figure~\ref{fig:allrf}).  This attenuated the coaxial
transmission of RF between the shields without significantly
compromising the thermal isolation between the different temperature
stages inside the cryostat.

Each individual bolometer was enclosed in a RFI-tight cavity.  The
optical opening into the cavity was small enough to prevent RFI at
frequencies below 100~GHz from radiatively coupling onto the
bolometers.  The bolometer signals were filtered at the cavity wall by
LC feed-through filters.  Calculations of the high frequency
performance of these filters including distributed reactances were
made by Hristov (1999), private communication.

These measures were only partially implemented before \maximai.  The
aluminized mylar septa and bolometer cavity LC filters were
incorporated after \maximai\ and before \maximaii.  Tests revealed
that detectors with LC filters showed dramatic improvement.  The LC
filters appeared to suppress pickup at $1 < f < 2$~GHz in both
optical and dark detectors.  We determined that the septa and LC
circuits contributed at least 20-dB suppression of pickup in this
frequency regime.  The LC circuits were shown not to introduce
measurable suppression or phase shifts in the signal band of the
detectors.

This evidence supports the model that during \maximai, RFI leaked in
through the window, around the temperature baffles, and then coupled
into the bolometer wiring.  This is believed to have caused the small,
scan-synchronous signal seen by some \maximai\ detectors
(\S\ref{bolonoisechar}).

\subsection{Bias and Readout Electronics} \label{readout}

The analog bolometer readout electronics were designed by V. Hristov
from CalTech.  The bolometers were AC-biased to minimize low frequency
noise contributions from the electronics (e.g. 1/f noise in the JFET
amplifiers).  AC bias does not cause fluctuations in the
bolometer temperature, assuming constant optical power and $f_{bias}
>> 1/2\pi \tau_{bolo}$.  All detectors were biased at the same
frequency by a single sine wave oscillator which could be tuned from 250~Hz
to 400~Hz.  We set this frequency before each flight to minimize
microphonic noise.

The bias voltage amplitude for each detector was set between 50~mV and
100~mV by a potentiometer.  Bias current variations were adequately
controlled by two 40 $M\Omega$ load resistors in series with each
bolometer.  The largest fractional change in bias current, observed
during the
\maximai\ planet scan, was of order $10^{-5}$.

We minimized the microphonic sensitivity of the receiver by reducing
the output impedance of the bolometers.  This was achieved with a
matched pair of cooled monolithic-dual silicon EJ-TIA JFET amplifiers
(Infrared Laboratories (U.S.A.)) in a matched emitter follower circuit.
The JFET pairs were packaged in standard transistor cans with internal
thermal isolation and were cooled by a weak thermal link to the liquid \4he
bath, for an operating temperature of 150~K.  Each JFET pair consumed
300~\micro~W of power and contributed 6-10\nvrthz to
the total noise per channel.

The bolometer signals were further processed by ambient readout
electronics.  They were pre-amplified with an AD624 op amp, band-pass
filtered around the bias frequency, and then were each rectified with an
AD630 lock-in amplifier referenced to the bias signal.

Bolometer resistance fluctuations appeared in the side bands of the
bias frequency, as determined by the telescope scan speed.  Given the
primary mirror modulation frequency of 0.45 Hz and peak velocity of
4~deg~sec$^{-1}$ (or \sima 25 beam FWHM per second), the data were in the
range of 0.1-20~Hz around the bias frequency.  Signals outside the
band were suppressed by a four pole Butterworth filter with a
characteristic frequency of 19.96\err 0.61~Hz, averaged over all 20
read-out channels.

Finally, the rectified signal was amplified by a gain of 1800 and
split to two outputs.  We removed the offset from one signal stream
with a single pole high-pass filter with a characteristic frequency of
14.9\err 0.7~mHz, averaged over all 20 read-out channels. This signal
was further amplified by a factor of 36, yielding a net gain of
65,000.  Overall, the gain of the readout circuit had a temperature
stability of $< 10^{-5}$ (\dega~C)$^{-1}$.
	
\subsection{Supporting Electronics and Telemetry}

Bolometer signals were amplified, rectified, and digitally sampled
with 16-bit resolution every 4.8~ms.  Other data, including cryostat
housekeeping, ambient temperatures, bolometer bias monitors, primary
mirror position, electronics status, and low gain bolometer signals
were sampled every 19.2~ms.  The data were multiplexed and transmitted
to a ground based data recording station over a 1.8-GHz radio downlink
at a data rate of 160~kbits~sec$^{-1}$.  A separate 1.5-GHz downlink
transmitted for the data from the pointing system and from balloon
facility support systems.  Video signals from pointing sensors were
transmitted at 2-3~GHz, but were not transmitted during CMB scans to
minimize the ambient radio frequency radiation.  Control of the
telescope from the ground was maintained via intermittent command
signal at \sima 500~MHz.

The digitization and multiplexing data acquisition system was built at
the Lawrence Berkeley National Laboratory, while telemetry and
commanding systems were provided by the National Scientific Balloon
Facility.

\section{CALIBRATION}\label{calibration}

		Absolute responsivity calibration used two known
sources during each \maxima\ flight: the CMB dipole and a planet.
Measurements of the CMB dipole gave the best absolute calibration for
the 150-GHz and 240-GHz detectors.  Observations of planets (Jupiter
in \maximai\ and Mars in \maximaii) were used to calibrate the 410-GHz
detectors, and were used as a consistency check for the dipole
calibration.  The internal millimeter wave source (the stimulator
described in \S\ref{receiv:stim}) was used to periodically measure
responsivity changes.

		The \maximai\ data have a calibration error of 4$\%$,
while the \maximaii\ data have a calibration error of 3$\%$.  These
are the most accurate calibrations achieved by any sub-orbital CMB
experiment.  The \maxima\ calibrations are consistent with that of
\wmap\ \citep{abroe_crosscor}.

\subsubsection{Linearity}

	All calibration sources used for \maxima\ are significantly
brighter than CMB fluctuations.  Planet calibrations and stimulator
events added enough loading to measurably reduce detector responsivity.
These linearity changes are summarized in Table~\ref{table:linearity},
and their impact on the calibrations is discussed in
\S\ref{calib:diperror} (dipole), \S\ref{calib:planerror} (planets),
and \S\ref{calib:stimerror} (stimulator).

\begin{table}
\caption{Calibration Linearity}
\begin{center}
\begin{tabular}{ccc}
\hline
 & Planet & Stimulator
\\ & Responsivity & Responsivity
\\ & Change & Change
\\ & \tiny{(\maximai/\maximaii)} & \tiny{(\maximai/\maximaii)}
\\\hline
 150~GHz & \small{0.7-2.5\%/0.2-0.5\%} & \small{0.1-0.5\%/0.1-1.5\%}
\\ 240~GHz & \small{1.0-6.1\%/0.2-0.4\%} & \small{0.2-4.0\%/0.1-0.5\%}
\\ 410~GHz & \small{1.9-7.7\%/0.4-1.0\%} & \small{0.5-3.3\%/0.3-2.7\%}
\\\hline
\multicolumn{3}{l}{Note.- Quoted values were derived from the maximum}\\
\multicolumn{3}{l}{of the signal for \maximai\ (before the slash) and}\\
\multicolumn{3}{l}{\maximaii\ (after the slash).  Ranges represent variations}\\
\multicolumn{3}{l}{between detectors.  For the dipole calibration the}\\
\multicolumn{3}{l}{upper limit on nonlinearity was \sima 0.05\%.}
\end{tabular}
\label{table:linearity}
\end{center}
\end{table}

\subsection{CMB Dipole}\label{calib:dipole}

	The dipole was the main calibrator for the 150-GHz and 240-GHz
detectors, using the amplitude measurement of the COBE satellite
\citep{smootdipole}. Dipole observations were carried out by rotating
the telescope in azimuth with an azimuthal angular velocity of
20\dega~sec$^{-1}$.  The observed signal from the dipole was a sine
wave at 56~mHz.  We observed the dipole at an elevation of 50\degs in
\maximai\ (to avoid side-lobe response from the Moon at lower
elevation) and at 32\degs in \maximaii.  During \maximai\ the dipole,
including the motion of the Earth around the Sun, had a magnitude of
3.195~mK and was tilted 20\degs from the horizon; the observed
amplitude over the scan region was 2.04~mK.  During \maximaii\ the
dipole had a magnitude of 3.010~mK and was tilted 48\degs from the
horizon; the observed amplitude was 1.15~mK.

	The rotating scan pattern of the dipole observation was
sensitive to parasitic signals from Galactic dust and in some cases
the atmosphere.  The Galactic dust signal was modeled from frequency
extrapolations of published maps (\citet{forecast}, \citet{sfd_dust}).
The dust signal was much smaller than the dipole signal at 150~GHz,
except near the Galactic plane.  Data within 5\degs of the Galactic
plane were neglected in data analysis.  Elsewhere, the dust model was
fit to the data along with the dipole model.  Overall normalization was
taken as a free parameter to account for uncertainties in the
frequency extrapolation of the dust signal.  In practice, the dust
model did not affect the dipole calibration due to its low amplitude
and lack of a dipole-like spatial component.

	An additional small signal was observed in the beginning of
the \maximai\ dipole calibration.  In \maximai\ we began the dipole
observation near the beginning of the flight, while the telescope was
still ascending from \sima 21.5~km to the final observing altitude of
\sima 38.5~km.  The additional signal was observed during the first
third of the observation (altitude $<$30~km).  We believe that this
signal was atmospheric for four reasons: (1)~it was highly correlated
in all the optical bolometers; (2)~it was spectrally consistent with
atmospheric emission, being larger for the higher frequency detectors;
(3)~it was spatially stable on the scale of a few minutes, but varies
on longer scales; (4)~the magnitude of the signal declined steadily
with altitude.

	The atmospheric signal was corrected using data from the
410-GHz bolometers.  These data, which were relatively insensitive to
the dipole and sensitive to the atmospheric signal, were used as a
template for the signal in the 150-GHz and 240-GHz data.  A correction
was applied for the CMB sensitivity of the 410-GHz detectors, as
calibrated by planet observations.  As with Galactic dust, we found
that fitting the atmospheric signal did not affect our final dipole
calibration values.  It did, however, increase calibration
uncertainty, because of noise in the 410-GHz data used to model the
effect.

\subsubsection{Dipole Data Analysis}\label{calib:dipdata}

	During each flight, the dipole was observed for \sima 30
minutes (100 rotations).  For each detector, the effects of electronic
filters and bolometer time constants were first deconvolved from the
entire data stream.  Data from each rotation were then fit
independently according to the model,
\begin{eqnarray}
T_{detector} = && (A * T_{CMB,Model}) + (B * T_{dust}) + \nonumber\\ && (C * N_{Drift}) + D
,\end{eqnarray}
\noindent in which $T_{detector}$ is time stream of detector data in
voltage units, $T_{CMB,Model}$ is the CMB dipole model in units of
temperature contrast, $T_{dust}$ is the Galactic dust model,
$N_{Drift}$ is linear drift, and $A$, $B$, $C$, $D$ are fitting
constants.  $A$ is the calibration of the detector to CMB signals.

	For \maximai\ the atmospheric signal was taken into account by modifying the fit to,
\begin{eqnarray}\label{eqn:dipmodelfull}
T_{detector} = && (A * T_{CMB,Model}) + (B * T_{dust}) + \nonumber\\ && (C * N_{Drift}) + D + (E * T_{410})
.\end{eqnarray}
\noindent $T_{410}$ is time stream data from a 410-GHz detector and
E is an additional fitting parameter.  Because the 410-GHz data do
have some very small sensitivity to the CMB, $A$ is no longer an
unbiased calibration.  To account for this, the 410-GHz data are first
calibrated using planet data and the parameter $A$ is
corrected,
\begin{equation}
A' = A - (E * Cal_{410}).\end{equation}
\noindent $Cal_{410}$ is the planet-based calibration of the 410-GHz
data and $A'$ is the true calibration of the low frequency channel.
In practice the 410-GHz term does not affect calibration values by more
than 0.5~$\sigma$, though it does increase uncertainties.  The correction
from $A$ to $A'$ has a negligible effect on both the calibration and
the calibration uncertainty due to the small value of the $E$ parameter.

	Each rotation yielded a calibration value ($A$ or $A'$ above)
and an associated error range.  These were combined statistically, with
2-$\sigma$ outliers excluded.  Data from \sima 80 rotations were analyzed
from each flight, with 2 to 8 excluded as outliers for each detector.

\begin{figure}[ht]
\plotone{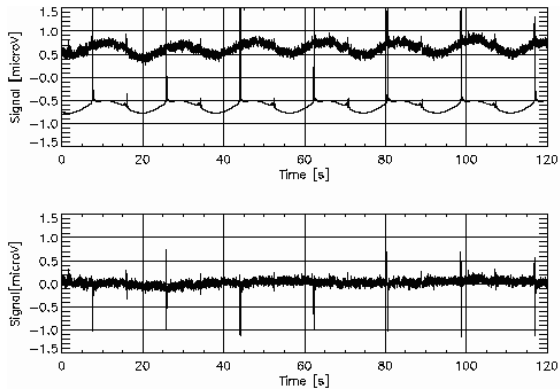}
\caption{\maximaii\ 150-GHz dipole data and
fit.  {\bf{Top panel:}} The top trace is the data from a 150-GHz
bolometer during observations of the CMB dipole.  An overall gradient
is removed and the offset is arbitrary.  The sinusoidal signal
is the CMB dipole modulated by the rotation of the telescope (\sima 18
second period).  The large spikes are caused by dust near the Galactic
plane.  The lower trace is a fitted model curve, including the CMB
dipole and a Galactic dust map.  {\bf{Bottom panel:}} The difference
between the model and the fit in the top panel are shown.  The model
deviates from the data near the Galactic plane crossing which are not
well fit with the \sima 10\mins pointing reconstruction accuracy of
the dipole observation.}
\label{fig:dipole}
\end{figure}

	Data from the \maximaii\ dipole observation are shown in
Figure~\ref{fig:dipole}.

\subsubsection{Dipole Calibration Error Sources}\label{calib:diperror}

	Dipole calibration uncertainty (1-4\% depending on the
detector) was dominated by detector noise at the dipole observation
frequency of 56~mHz.  The fitting routine was found numerically to
reject noise beyond a fractional bandwidth of \sima 0.5.  Noise was
effectively reduced by a further factor of $\sqrt{2}$ by the known
phase of the dipole model.  Raw detector noise near 56~mHz for a
150-GHz bolometer was typically 150~nV~Hz$^{-0.5}$.  Considering
bandwidth and phase constraints, this yielded an expected \sima 20-nV
noise level for a single dipole fit.  For a typical 150-GHz detector
the amplitude of the dipole response was \sima 70~nV.  We therefore
expected a statistical uncertainty of \sima 30\% from a single
rotation.

	Detector noise was the only source of statistical uncertainty
in the calibration, and could be estimated directly from the scatter
of the individual, single rotation calibrations.  Such analysis yielded
single rotation statistical uncertainties of 10\% to 30\% for 150-GHz
detectors in either flight.  These numbers were somewhat lower than
predicted due to imperfect understanding of detector noise at very low
frequencies.

	An integration of 80 to 90 dipole observations per flight
provided a total statistical uncertainty of 1.4\% to 4.2\% for 150-GHz
detectors in \maximai\ and 1.1\% to 2.5\% in \maximaii.  Because the
CMB responsivity of the 240-GHz detectors was 60-70\% that of the
150-GHz detectors, they had a proportionally higher statistical
error.  When combining data from multiple detectors, we used the
highest statistical uncertainty of the combined channels.  For
\maximai, the modelling of the atmospheric signal increased the final
calibration uncertainty by only about 0.5\%.

	In addition to statistical uncertainty, there were a number of
known systematic effects, though none had a significant impact on the
calibration.  First, dipole pointing reconstruction was accurate to
\sima 10\mins and contributed negligibly (\sima 0.1\%) to the
calibration uncertainty.  Second, the signal from the dipole was small
enough that bolometer saturation was negligible.  Third, the dipole
model derived from the COBE measurement was accurate to 0.68\%.
Finally, a \sima 15~mHz single-pole high-pass filter in the bolometer
readout electronics was characterized to \sima 0.3\% at the dipole
scan frequency.  Though this was tested under laboratory conditions,
where the electronics were substantially cooler than in flight, the
temperature coefficients of the filter caused at most a 1\% shift in
filter pole frequency in flight.  This maximum shift would have caused
only a 0.08\% change in response magnitude and 0.16\degs change in
phase.

\subsection{Planets}\label{calib:planet}

		In each flight observations were made of a planet:
Jupiter in \maximai\ and Mars in \maximaii.  The planet observation
procedure has been described in \S\ref{planet}.  Responsivity
calibration was obtained from the maximum voltage response.  Expected
signals were derived from published measurements and models of planet
temperatures and emissivities (\citet{jupiter}, \citet{mars1},
\citet{mars2}), combined with the spectral response of the detectors.
A correction was applied for beam dilution, i.e. the fraction of the
telescope beam filled by the planet.  Jupiter had an angular diameter
of 46.5\asec\ during \maximai\ and Mars had an angular diameter of
12.7\asec\ during \maximaii.  The dilution factors varied from
3.4$\times 10^{-3}$ to 4.4$\times 10^{-3}$ for \maximai\ and
3.1$\times 10^{-4}$ to 4.1$\times 10^{-4}$ for \maximaii.  This
correction is the dominant error source for the planet calibration in
both flights.

		An additional correction of roughly 5\% for \maximai\
and 1\% for \maximaii\ was applied for the reduction in responsivity
caused by the optical load from the planet
(Table~\ref{table:linearity}).  This effect was neglected in the
initial \maximai\ data analysis and caused an apparent small
systematic discrepancy between the dipole and Jupiter calibrations.
This discrepancy was within the error range of the Jupiter calibration
and did not affect the CMB map or power spectrum.

\subsubsection{Planet Calibration Error Sources}\label{calib:planerror}

		The dominant error term for the planet calibration is
the uncertainty in the beam dilution factor.  The uncertainty in the
integrated beam response is 5\% to 10\%.  In addition, there is a
possibility of small, broad side-lobes that are not measured in the
beam maps.  We assign an uncertainty of 10\% from beam shape errors.
Beam shape error, especially that due to broad side-lobes, is partially
correlated between detectors because of their shared optics.

		Uncertainties in the effective brightness temperature
of the planets contribute 5\% to calibration error.  The brightness
temperature of Mars has been modeled to this accuracy, both by
extrapolation from high frequency observations \citep{mars1} and by
physical modeling \citet{mars2}.  The atmospheric properties for
Jupiter make modeling relatively difficult.  Our expected Jupiter
signal is based on published brightness ratios between Jupiter and
Mars \citep{jupiter}.  The planet temperature uncertainty is fully
correlated between all the detectors.

		Measurements of the detector spectra contribute 1-2\%
error at 150~GHz, 3-7\% at 240~GHz, and 2-3\% at 410~GHz.
Measurements of the peak planet voltage contribute 1-4\% error; one
detector in \maximaii\ is anomalously noisy, increasing this term to
\sima 10\%.  Uncertainty in the bolometer saturation is negligible.

\subsection{Time Dependent Calibration}\label{calib:relative}

		The responsivity of bolometers varied by 1-2$\%$ for a
change of 1~mK in the thermal reservoir temperature.  The temperature
varied by \sima 6~mK over the course of data collection in
\maximai\ and by \sima 21~mK in \maximaii.

		Responsivity variations were monitored using the
internal millimeter wave source (stimulator) described in
Section~\ref{receiv:stim}.

		To obtain relative calibration values from stimulator
events, we began by subtracting an overall gradient from each event to
remove the effects of detector drift.  We then performed a linear fit
between pairs of stimulator events.  The slope of this fit was the
calibration ratio between the events, while the offset of the fit was
simply an offset in the detector data.

\begin{figure}[ht]
\includegraphics[angle=90,width=3.0in]{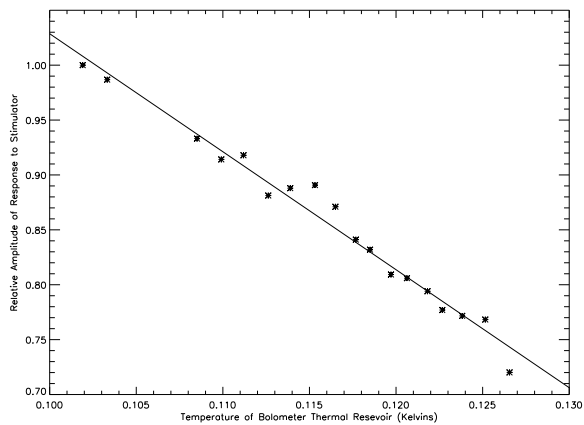}
\caption{Temperature dependence of the responsivity of a 150-GHz detector.  These data were
collected during \maximaii.  The point near 0.127~K, measured shortly
after sunrise, showed less responsivity than would be expected from
the nighttime data.  During \maximai, the temperature of the thermal
reservoir was more stable, varying from 98-104~mK.}
\label{fig:relative_cal}
\end{figure}

		Once the relative calibration at each stimulator event
was known, we fit the values to a linear function of the temperature of
the bolometer thermal reservoir (Figure~\ref{fig:relative_cal}).  This
fit was combined with the absolute calibration to obtain the overall
calibration as a function of time throughout the flight (\S\ref{calib:final}).

\subsubsection{Relative Calibration Error Analysis}\label{calib:stimerror}

		The relative calibration between stimulator events was
affected by random variations (detector noise or stimulator
instability), but was not affected by systematics that are consistent
between stimulator events, such as uncertainties in the spectra of the
detectors and the beam filling of the stimulator signal.  The
reduction in bolometer responsivity due to the large optical load of
the stimulator was nearly stable (Table~\ref{table:linearity}).  Though
it did vary with bolometer temperature, this variation contributed
less than 0.1\% error to the relative calibration.

		Random errors in the comparison of stimulator events
were 1-2\%.  Instabilities in the stimulator current accounted for
$<$0.5\% of this, while detector noise accounts for the rest.

	We treated the relative calibration as a linear function of
temperature and took the optical load during the flight as constant.
These assumptions were supported by bolometer models
(\citet{CDWThesis}, \citet{grannan_bolo}) and contributed negligibly
to calibration error.

\subsection{Combined Calibration}\label{calib:final}

		The overall calibration for each detector was obtained
by combining an absolute calibrator with the relative calibration.
The absolute calibrator was the CMB dipole for 150-GHz and 240-GHz
detectors and was the planet scan for 410-GHz detectors.  The relative
calibration was based on the temperature of the bolometer thermal
reservoir and the responsivity-temperature relation obtained in
Section~\ref{calib:relative}.  Temperature was monitored continuously.

		The overall calibration error was the combined error
from the absolute calibrator and the relative calibration.  Relative
calibration error varied over the course of the flight.  Quoted values
are based on averages over the CMB observations.  Relative calibration
error was subdominant for \maximaii\ (1\% to 2\%) and was negligible for
\maximai\ ($<$0.1\%).

		The published \maximai\ data were conservatively
assigned the highest calibration uncertainty of the detectors used,
4\%.

\section{SCANS AND POINTING}\label{pointing}

\subsection{Scan Strategy}\label{point:scans}

		Each CMB observation was conducted at a fixed
elevation, while the telescope beams were moved in azimuth.  The
azimuth modulation defined one dimension of the roughly rectangular
scan region.  The rotation of the sky over the duration of the
observation defined the other dimension.

		The azimuth motion consisted of two independent
modulations.  The primary mirror rotation provided a relatively fast
modulation with a frequency of 0.45~Hz and a peak-peak span of
4.0\dega .  The motion of the entire telescope provided slower
modulation at frequencies from 12~mHz to 25~mHz, depending on the scan
parameters, with peak-peak spans of 4.5\degs to 9.0\dega .  The fast
modulation prevented our data from being significantly corrupted by
low frequency noise.  Taken together, the two made our data relatively
robust against modulation synchronous parasitic signals.

		The two CMB scans of each flight observed the same
region of the sky, but at different times and at different elevation
angles.  The rotation of the sky between these observations caused the
two scans to be tilted relative to each other in sky-stationary
coordinates.  The average angle of cross-linking was 22\degs for
\maximai\ and 27\degs for \maximaii.

		The two azimuth modulations, the rotation of the sky,
and the cross-linked revisitation were uncorrelated and are on radically
different time scales, minimizing the effects of potential
scan-correlated systematics.

		Depth of integration was set by the total width of the
combined azimuth modulation and the rotation rate of the sky.  The
average integration time per beam-size was \sima 2.5~seconds in
\maximai\ and \sima 2.2~seconds in \maximaii.  This led to an
average expected noise level of \sima 60~\micro K per beam size area
for our best single detector and \sima 40~\micro K for our published
combination of four \maximai\ detectors.  In practice, integration was
several times longer in the center of the observed region and shorter
near the edges.

\subsection{The Attitude Control System}\label{point:hardware}

		The pointing system, or attitude control system (ACS),
served both to control the orientation of the telescope in flight and
to acquire the data needed for post-flight pointing reconstruction.
The pointing system consisted of attitude sensors, a central feedback
loop control computer, and motors.  Some of the easily interpreted
sensors were used in pointing control, while the most precise sensors,
the CCD cameras, were used after the flight for pointing
reconstruction.  The ACS was originally constructed at IROE-CNR (now
IFAC-CNR) and Universit\'a di Roma La Sapienza, in Italy.  The system
was similar to that used in the BOOMERanG experiment \citep{boom}.

		A computer read data from the various sensors, applied
a digital feedback algorithm, and set the power level for the motors.
Each of these tasks was performed once every 96~ms, synchronously with
the bolometer data acquisition system.

		Signals from most sensors were sampled each cycle.
Data from the CCD cameras were processed by a separate computer, and
were passed to the control computer every two cycles (192~ms).  GPS
data (absolute time and position) were updated once per second.

	Pointing normally followed one of several preprogrammed flight
schedules.  Remote (ground-based) commanding was used to switch
between, modify, or override schedules.  In addition, remote
commanding was used to modify control loop gains, to make
adjustments to sensor calibrations, and to set parameters for the CCD
image processing.

		Feedback control was based on azimuthal rotation
velocity, as measured by a rate gyroscope.  The gyroscope had an
accuracy of \sima 0.01\dega~sec$^{-1}$.  Though they were very sensitive, the
gyroscopes had substantial low frequency drifts, primarily due to
ambient temperature fluctuations.  Drifts were calibrated once per
gondola scan period, and had little impact on pointing control.  Two
other gyroscopes, measuring pitch and roll velocities, were not used in
feedback control.  The performance of the feedback control loop is
summarized in Table~\ref{tab:point_perform}.

\begin{table}
\caption{Pointing Performance}
\begin{center}
\begin{tabular}{ccccc}
\hline
CMB & \maximai & \maximai & \maximaii & \maximaii
\\Observation & Scan 1 & Scan 2 & Scan 1 & Scan 2
\\\hline
Max Scan Speed (deg~sec$^{-1}$)&0.29& 0.30& 0.29& 0.26
\\RMS Velocity Error (\ditto)& 0.022 & 0.028& 0.043 & 0.057
\\\hline
\end{tabular}
\label{tab:point_perform}
\end{center}
\end{table}

		Absolute azimuth was measured in real time using a two
axis magnetometer.  The magnetometer was extremely precise
($<$0.5\mina) in differential measurement, but was highly non-linear
due to the magnetic properties of the telescope.  Pre-flight
measurements were used to calibrate the magnetometer to an accuracy of
\sima 30\mina.

		Absolute elevation was measured in real time by an
optical angle encoder between the inner assembly (receiver and primary
mirror) and the outer frame of the telescope.  The accuracy of this
measurement depended on the balancing of the telescope (\sima
0.1\dega) and on long time scale pendulum motion (\sima 0.5\dega,
varying over tens of minutes).  The differential accuracy of this
elevation measurement was \sima 1\mina.

\subsubsection{CCD Cameras}

		The CCD star cameras provided the most accurate
measurement of telescope orientation and were used for post-flight
reconstruction.  One camera was mounted on the inner telescope
assembly and was boresighted with the telescope beams when the primary
mirror was in its central position.  The second camera was positioned
on the outer frame so that north celestial pole star (Polaris) was in
the camera's field of view during CMB observations - approximately
40\degs to the right of the boresight, at a fixed elevation angle of
about 31\dega .

		The boresighted camera data were used for the final
pointing reconstruction and were accurate to \sima 0.5\mina.  The
secondary camera was only accurate to \sima 15\mina, due to the offset
of the measurement from the telescope boresight combined with pendulum
motion of the telescope.  Secondary camera data were used to identify
stars in the primary camera.

	The boresighted camera had a field of view of 7.17\degs by
5.50\degs with a pixel size of 0.84\mins by 0.69\mina.  The resolution
was further improved to \sima 0.5\mins in flight by software
interpolation.  The secondary camera had a larger field of view
(14.34\degs by 11.00\dega) and lower resolution (\sima 1.0\mina).

	The boresighted camera reliably detected stars of V magnitude
5.0 or brighter.  Stars of V magnitude 5.0 to 6.0 were detected
intermittently, and stars dimmer than V magnitude 6.0 were rarely
detected.  This sensitivity was sufficient to detect stars in all of
the \maximai\ scan region and \sima 80\% of the \maximaii\ scan region.
Source brightness was not an issue for the secondary camera, which
always viewed Polaris (V magnitude = 2.0).

Image processing introduced a delay of \sima 200~ms to the camera
data.  The cameras internally sampled the CCD chips at 30~Hz,
asynchronous to the rest of the system.  This caused a jitter up to
33~msec in the image processing delay, for an overall delay of 200\err
16~ms between pointing data and bolometer data.  At our gondola scan
speeds, 200~ms translated to 2\mins to 5\mins of gondola rotation,
which is a significant fraction of our 10\mins beam-size and was
taken into account.  The \err 16-ms jitter translated to \sima
0.15\mins RMS pointing uncertainty.

\subsubsection{Motors}

\begin{figure}[ht]
\plotone{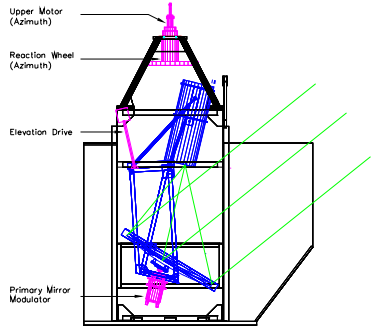}
\caption{Two motors near the top of the telescope
controlled azimuthal orientation by driving against a reaction wheel
and the cables from the balloon, respectively.  A linear
actuator/servo-arm tilted the inner assembly, pointing the telescope
in elevation.  A motor below the primary mirror modulated it at
relatively high speed (0.45~Hz, \err 2\degs amplitude) in azimuth.
The frames shown on the front and back of the telescope were far
side-lobe baffles.}
\label{fig:motors}
\end{figure}

		Three motors were used to point the \maxima\ telescope
and a fourth was used to modulate the primary mirror.  They are shown
schematically in Figure~\ref{fig:motors}.  Two motors, located near
the top of the telescope frame, were used for pointing in azimuth.
One of these drove a reaction wheel with a moment of inertia of
10~kg$\cdot$m$^2$(\sima 0.5\% that of the telescope).  The other
torqued against suspension cables connected to the balloon, which had
a much greater moment of inertia than the telescope.  Both motors were
direct drive (ungeared), had a peak torque of 33.9~N$\times$m, and
received a maximum power of \sima 50~W from the control system.  The
light reaction wheel provided fast response, while the other motor
kept the speed of the reaction wheel low by transferring angular
momentum into the balloon.  The rotational velocities of these motors
were monitored by tachometers.

		Elevation control was provided by a geared motor
connected to a linear actuator arm.  The arm was fixed between the
outer assembly of the telescope frame and the inner assembly of the
receiver and primary mirror.  The inner assembly was balanced about the
rotation axis, so the static load on this motor was very small.

		Motor power was determined in a digital control loop.
Pointing control in azimuth and elevation were not strongly coupled and
may be considered separately.

		In azimuth, we used a rotational velocity-based
proportional feedback system based on the rotational velocity measured
by a rate gyroscope.  In CMB scans, the target velocity was constant,
except during turnarounds in the scan direction.  During turnarounds
the target velocity varied linearly with time.  The absolute position
was not used directly in the control loop; it was instead used to
trigger these turnarounds.  An additional feedback term proportional
to the rotational velocity of the flywheel was applied to the upper
motor, slowly transferring flywheel angular momentum to the balloon.

The elevation control formula was based on the measured angle of the
telescope inner assembly relative to the outer frame.  In this case,
the power to the elevation drive was determined by a position-based
proportion-derivative feedback loop.

\begin{figure}[ht]
\plotone{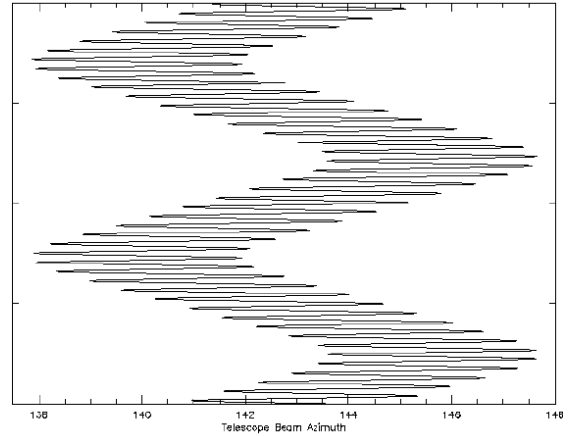}
\caption{A simulation of the
double modulation in azimuth.  The x-axis is the azimuthal position of
the telescope beams, while the y-axis is time.  The slower modulation
is caused by the motion of the entire telescope, while the faster
modulation is caused by the rotation of the primary mirror about the
optic axis.}
\label{fig:doublemod1}
\end{figure}

\begin{figure}[ht]
\includegraphics[angle=90,width=3.0in]{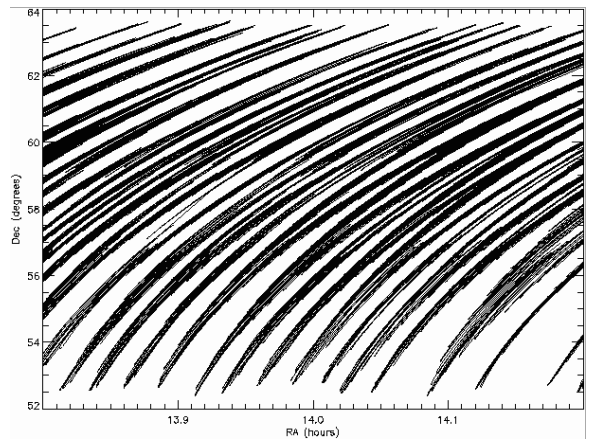}
\caption{A \maximaii\ scan pattern formed in RA and declination, combining the
azimuth modulations with the rotation of the sky.  Lines of constant
elevation move with the rotation of the sky, spanning the plot in a
diagonal arc from the lower left to the upper right.  The gaps seen in
this scan pattern are less than half the telescope beam size.}
\label{fig:doublemod2}
\end{figure}

\begin{figure}[ht]
\includegraphics[angle=90,width=3.0in]{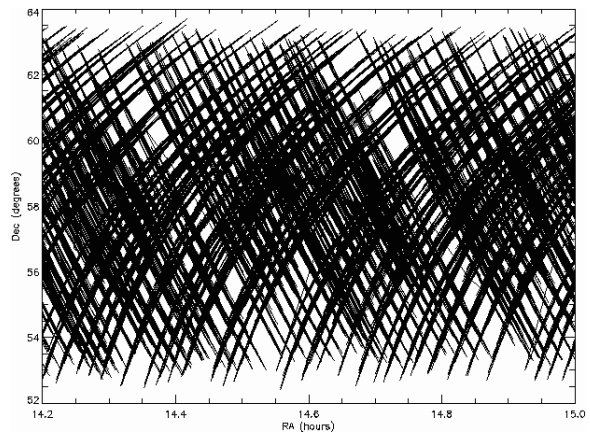}
\caption{The cross-linked scan pattern
from \maximaii, consisting of two scans similar to that shown in
Figure~\ref{fig:doublemod2}.  The average cross-linking angle was
27\dega.}
\label{fig:doublemod3}
\end{figure}

	The primary mirror was continuously rotated from side to side
about the axis indicated in Figure~\ref{fig:optics} in
Section~\ref{optics}.  The motion was a rounded triangle scan with an
amplitude of \err 2\degs and a frequency of 0.45~Hz.  This modulation
superimposed on that of the entire telescope yielded the scan pattern
in Figure~\ref{fig:doublemod1}.  The resulting scan pattern in RA and
Dec is show in Figures~\ref{fig:doublemod2} and~\ref{fig:doublemod3}.


	Because the mirror rotation axis was not vertical, there was
also a small modulation in elevation - a bowed pattern in which the
extremes of the mirror modulation rose slightly in elevation.  The
elevation motion could in principle have led to a scan synchronous signal
in the bolometers due to atmospheric emission.  In practice, the
elevation motion ($<$2\mina) was not large enough to generate a
detectable signal.

	The mirror was actuated by a DC motor with solid state PID
control electronics.  The motor was obtained commercially and the
control electronics were built at Berkeley.  Mirror position was
controlled to an accuracy of 1\mina .

\subsection{Pointing Reconstruction}\label{point:data}

	The complete pointing reconstruction for one detector for both
\maxima\ flights can be seen in Figure~\ref{fig:colorscans}.

	Pointing reconstruction was based primarily on data from the
CCD cameras.  Often, pointing could be determined simultaneously from
two different stars in the same field of view.  In these cases the
discrepancy between the two pointing solutions was used to estimate the
overall error in the CCD measurement.  The typical discrepancy was
\sima 0.5\mins (Table~\ref{tab:pointing_error}).

\begin{table}
\caption{Pointing Reconstruction Uncertainties}
\begin{center}
\begin{tabular}{cccccc}
\hline & \multicolumn{3}{|c|}{\bf{Random Errors}} & \multicolumn{2}{|c|}{\bf{Systematic Errors}}
\\	 		& CCD 		& Camera 	& Interpolation$^{c}$	& Detector	 	& Primary
\\	 		& Camera$^{a}$ 	& Timing$^{b}$	&		& Offset$^{d}$		& Modulation$^{e}$
\\\hline
\maximai 		& 0.46\mina & 0.15\mina & $<$0.001\mina & 0.25\mina & 0.81\mina
\\\maximaii 		& 0.58\mina & 0.15\mina & $<$0.001\mina (\sima 1\mina) & 0.25\mina & 0.81\mina
\\\hline
\multicolumn{6}{l}{$^{a}$CCD Camera-Error in image analysis.}\\
\multicolumn{6}{l}{$^{b}$Camera Timing- Uncertainty in time of image acquisition.}\\
\multicolumn{6}{l}{is the effect of the timing uncertainty of image acquisition.}\\
\multicolumn{6}{l}{$^{c}$Interpolation-$<$0.001\mina interpolation error applies to all of}\\
\multicolumn{6}{l}{\maximai\ and \sima 80\% of \maximaii.  The rest of \maximaii\ had}\\
\multicolumn{6}{l}{very few stars, increasing this term.}\\
\multicolumn{6}{l}{$^{d}$Detector offset-Boresight uncertainty.}\\
\multicolumn{6}{l}{$^{e}$Primary Modulation-Uncertainty in mirror orientation.}
\end{tabular}
\label{tab:pointing_error}
\end{center}
\end{table}

	Between camera measurements, the telescope typically moved
\sima 8\mina.  In the azimuth, data from the rate
gyroscope were integrated to find intervening positions.  As a test of
accuracy, the azimuth data were reinterpolated numerically and were
compared to the gyroscope-based interpolation.  The difference between
the two was used to estimate the pointing error introduced by
interpolation.  Though the RMS discrepancy between these two methods
was very small, the distribution had extreme outliers corresponding to
regions of the sky with few bright stars.  Interpolated regions with a
difference of greater than 3.3\mins (one third of the FWHM beam size)
were not used in data analysis.

	In the elevation direction, the rate gyroscopes were more
difficult to calibrate.  However, the motion of the telescope in
elevation was extremely slow and small, so these data were safely
interpolated numerically.

	Finally, the effect of the primary mirror modulation was
included.  The angle of the primary mirror was measured to several
arcseconds by a linear variable differential transformer (LVDT).  The
LVDT data were calibrated both before and during flight using data
from the planet observation.

		The motion of the primary mirror moved the telescope
beams primarily in azimuth.  However, there was a small motion in
elevation which was much more difficult to calibrate and was a source
of pointing uncertainty.  The elevation motion depended upon the zero
position of the mirror.  This was measured to \sima 1\dega, which led
to a conservative pointing uncertainty of about 0.8\mins RMS.

		The 0.8\mins uncertainty of the primary mirror
modulation was the largest source of error in the \maxima\ pointing
solution.  Though purely systematic, the scan pattern and
cross-linking tended to blur out the effect.  In addition, other
sources of pointing uncertainty further randomized the total error.
The overall pointing error was approximated as a 1\mins Gaussian blur.
End to end tests demonstrated that such a blur in leads to a 10\%
reduction in the CMB power spectrum at \ella =1000 and that this
reduction scaled roughly as \ella\sqra, over the range where the
fractional reduction is much less than 1.

		Pointing uncertainty was a subdominant source of CMB
power spectrum error at all values of \ella .  While it was possible to
compensate the power spectrum for the reduction caused by pointing
error, we have not done so because it was relatively small, and because
our model of the pointing error as Gaussian is not exact.

\section{SUMMARY AND DISCUSSION}\label{conclusion}

The performance of \maximai\ provided a definitive measurement of CMB
temperature anisotropy on sub-degree scales.  High resolution maps,
power spectra over multipoles of $35 \leq \ell \leq 1235$, and
cosmological parameter estimates were generated from data from five of
the sixteen photometers, as published in \citet{hanany_results},
\citet{balbi_results}, \citet{lee_results}, \citet{stompor_results}, and
\citet{abroe_results}.  Both intermediary and final results passed a
series of systematic tests (\citet{radek_cras}, \citet{abroe_crosscor}),
probing contamination from instrumental noise, from foregrounds or
from the data analysis pipeline.

The 225 square degrees area of the sky that was scanned during the
\maxima-2 flight in 1999 overlapped with 50 square degrees of the
area scanned during \maxima-1 and was larger by about a factor of two,
see Figure~\ref{fig:colorscans}.  The expected detector performance
and scan strategy were similar between the two flights.  However, the
data showed a somewhat higher level of systematic effects, which would
have required more effort to understand and overcome. The team has
decided to release only limited results that will facilitate the
comparison between the \maxima-1 and \maxima-2 maps
\citep{abroe_crosscor}.

\vskip -0.3cm
\acknowledgments 

JHPW and AHJ acknowledge support from NASA LTSA Grant no.\ NAG5-6552
and NSF KDI Grant no.\ 9872979.  PGF acknowledges support from the RS.
BR and CDW acknowledge support from NASA GSRP Grants no.\ S00-GSRP-032
and S00-GSRP-031.  MEA and RS acknowledge support from NASA grant
no. NRA-00-01-AISR-004.  \maxima\ is supported by NASA Grants
NAG5-3941, NAG5-4454, by the NSF through the Center for Particle
Astrophysics at UC Berkeley, NSF cooperative agreement AST-9120005.
Computing resources were provided by the National Energy Research
Scientific Computing center, which is supported by the Office of
Science of the U.S. Department of Energy under contract
no. DE-AC03-76SF00098, and by the University of Minnesota
Supercomputing Institute in Minneapolis, Minnesota.  \maxima\ field
and flight support was provided by the National Scientific Balloon
Facility.  The \maxima\ team would like to thank P. Timbie for the use
a Gunn oscillator.  We thank P. Mauskopf for providing electronic
readout units.


\newpage 




\end{document}